# Electronic Thermal Transport Measurement in Low-Dimensional Materials with Graphene Nonlocal Noise Thermometry

Jonah Waissman[1], Laurel E. Anderson[1], Artem V. Talanov[1,2], Zhongying Yan[1], Young J. Shin[1], Danial H. Najafabadi[1], Mehdi Rezaee[2], Xiaowen Feng[3], Daniel G. Nocera[3], Takashi Taniguchi[4], Kenji Watanabe[5], Brian Skinner[6], Konstantin A. Matveev[7], Philip Kim[1,2]*

[1]*Department of Physics, Harvard University, Cambridge, MA 02138, USA*

[2]*John A. Paulson School of Engineering and Applied Sciences, Harvard University, Cambridge, MA 02138, USA*

[3]*Department of Chemistry and Chemical Biology, Harvard University, Cambridge, MA 02138 USA*

[4]*International Center for Materials Nanoarchitectonics, National Institute for Materials Science, 1-1 Namiki, Tsukuba 305-0044, Japan*

[5]*Research Center for Functional Materials, National Institute for Materials Science, 1-1 Namiki, Tsukuba 305-0044, Japan*

[6]*Department of Physics, The Ohio State University, Columbus, OH 43210, USA*

[7]*Materials Science Division, Argonne National Laboratory, Argonne, IL 60439, USA*

**In low-dimensional systems, the combination of reduced dimensionality, strong interactions, and topology has led to a growing number of many-body quantum phenomena. Thermal transport, which is sensitive to all energy-carrying degrees of freedom, provides a discriminating probe of emergent excitations in quantum materials and devices. However, thermal transport measurements in low dimensions are dominated by the phonon contribution of the lattice, requiring an experimental approach to isolate the electronic thermal conductance. Here, we show how the measurement of nonlocal voltage fluctuations in a multiterminal device can reveal the electronic heat transported across a mesoscopic bridge made of low-dimensional materials. By using 2-dimensional graphene as a noise thermometer, we demonstrate quantitative electronic thermal conductance measurements of graphene and carbon nanotubes up to 70 K, achieving a precision of ~1% of the thermal conductance quantum at 5 K. Employing linear and nonlinear thermal transport, we observe signatures of long-range interaction-mediated energy transport in 1-dimensional electron systems, in agreement with a**

**theoretical model. Our versatile nonlocal noise thermometry allows new experiments probing energy transport in emergent states of matter and devices in low dimensions.**

Heat transport by electrons has been central to the study of materials ever since the pivotal measurements of Wiedemann and Franz (WF)[1]. For weakly-interacting electronic systems such as normal metals, the ground state is described by Landau's Fermi liquid paradigm[2], and electronic charge and heat flow are intimately connected, giving rise to the WF law. If interactions are sufficiently strong, weakly-interacting charged quasi-particles no longer describe system behavior and the WF law breaks down. Notable examples of such strongly interacting systems include quasi-1D materials[3], metallic ferromagnets[4], heavy fermion materials[5], underdoped cuprates[6], and the charge-neutral point of graphene[7], all cases related to the emergence of non-Fermi liquid behavior due to strong interactions.

In low dimensional systems, such as 2D quantum materials and 1D nanowires, these electronic interaction effects (including spin) are enhanced[8,9]. A growing number of strongly correlated states involving interactions and topology have been identified, including 1D and 2D electron Wigner crystals[10–12], strongly-correlated insulators and superconductors in twisted 2D heterostructures[13], and 2D magnets[14]. Thermal transport experiments are of immediate interest to clarify the nature of these materials[15]. Furthermore, the growing technological relevance of 2D and 1D materials demands an experimental probe of electronic energy transport and dissipation properties, separated from other degrees of freedom.

Accessing electronic thermal transport is challenging due to the prevailing phonon contribution[16]. In bulk materials, the electronic contribution can be extracted by extrapolation to the low temperature limit where the phonon contribution rapidly decays[6], by using a magnetic field to separate the electronic contribution with the thermal Hall effect[3], or by chemical doping to re-enter the WF regime and thereby estimate the phonon contribution in an isostructural sample[5]. Electronic thermal transport was also successfully isolated in some mesoscopic systems, notably the quantum Hall effect[17–19], single-electron transistors[20], and atomic contacts[21,22], by implementing electronic thermometry specific to the system of study[7,23–26]. However, for low-dimensional materials, such as 2D van der Waals monolayers[27] and 1D nanotubes[28] and molecular junctions[29,30], existing techniques are dominated by phonon transport, and a method that quantitatively isolates the electronic contribution is yet to be realized.

We approach this problem using Johnson-Nyquist noise, the fluctuations of voltage or current arising due to the finite temperature of electrical conductors[31,32]. Classically, for a resistor $R$, the voltage fluctuations are given by $\langle V^2 \rangle = 4 k_B T R \Delta f$ where $T$ is temperature, $R$ is the electrical resistance, and $\Delta f$ is the measurement frequency bandwidth. Johnson noise is independent of the material type, size, or shape, operating over a wide frequency band and temperature range, and is thus widely used in fundamental science and applications[33]. In two-terminal mesoscale samples, Johnson noise can be used to measure electronic thermal conductance using self-heating[23,24,34,35], in which Joule power dissipated in a resistor is balanced by energy loss channels, generating a measured temperature rise. Recently, this approach was used for graphene[23,24,34,35], where electronic diffusion cooling governs energy loss over a wide temperature range, allowing electronic thermal conductance to be measured to $T > 100$ K. However, because the device under test is simultaneously the thermometer, this approach is limited to diffusive conducting states with low energy loss to phonons and low contact resistance, restricting its use to graphene. A thermal transport measurement that applies to other materials and non-diffusive conduction requires a minimum of two temperature inputs to specify the temperature gradient driving energy flow. We thus require a multiterminal approach, in which the local temperature of two points along a device is measured by fluctuations of corresponding local resistors.

The multiterminal generalization of Johnson noise can be found by considering noise in diffusive multiterminal conductors. An example is shown in Fig.1a. A conducting system is connected to multiple leads held at a bath temperature, which may be grounded or floating. Current is injected through one of these leads, causing Joule heating of the system. In the limit that electrons generate a local temperature through strong equilibration, known as the hot electron regime, it was theoretically shown[36] that the noise power measured between any two terminals $n$ and $m$ is given by $S_{nm} = \int_{-\infty}^{\infty} dt \langle \delta I_n(t) \delta I_m(t) \rangle = \int d\mathbf{r}\, g_{nm}(\mathbf{r})\, T_e(\mathbf{r})$. Here, $\delta I_n$ is the fluctuation current, $T_e(\mathbf{r})$ is the local electronic temperature and $g_{nm}(r)$ is a geometry-dependent local weighting function defined as $g_{nm}(r) = \nabla\phi_n \cdot \hat{\sigma}\nabla\phi_m$, where $\hat{\sigma}$ is the local conductivity and $\phi_n$ is a characteristic potential associated with each terminal of the device (see Supplementary Section 4 for further details). This relationship holds if the energy supplied by Joule heating remains in the electronic degrees of freedom and energy losses to phonons and other heat sinks is negligible. Under these conditions, the noise emitted at any terminal is closely related to the energy transported to that region of the device and the resulting electronic temperature distribution.

For a thermal conductance measurement, we seek to realize the thermal circuit sketched in Fig.1b, in which two measured temperatures, $T_H = T_{bath} + \Delta T_H$ and $T_C = T_{bath} + \Delta T_C$, are combined with the energy current $Q$ across a bridge between two thermometers to give the two-terminal thermal conductance $G_{bridge}^{th} = \frac{Q}{T_H - T_C}$. To implement this with multiterminal noise, we utilize the geometry shown in the central panel of Fig.1c. The device possesses four terminals, divided into two pairs. Each pair contacts a rectangular conducting region, defining two diffusive electronic thermometers. A bridge connects the two rectangular thermometers at their midpoints and serves as the material of interest, which need not be diffusive. The wider rectangle on the left serves as the hot side where diffusive electrons are Joule heated by an injected low-frequency current. To avoid directly Joule heating the bridge and cold side, the heating circuit is balanced such that only energy current $Q$ and no electrical current traverses the bridge (see Supplementary Section 2 for details). The bridge width is narrow compared to the hot side length such that it obtains a thermal bias at the peak of the hot side temperature distribution. The narrow right-most rectangle serves as the cold side. Energy current across the bridge heats the cold side at its center point and is equilibrated at the cold contacts, generating a peaked temperature distribution and nonlocal voltage fluctuations. The narrow cold side design optimizes sensitivity by maintaining a maximal average, while the wide hot side ensures a local temperature distribution that is insensitive to the bridge. This geometry can be generalized to multi-thermometer setups such as for thermal Hall measurement by appending two-terminal thermometers at other points along the bridge (see Supplementary Section 9).

Each thermometer should measure the local temperature $T_{H,C}$ without cross-contamination of signals, despite being in electrical contact. In a two-terminal rectangular geometry with a uniform conductivity, the local weighting function $g_{nm}(r)$ is well-approximated as a constant and $S_{nm}$ is proportional to the average $T_e(r)$. To maintain this constant weighting in the multiterminal case, we define a differential noise correlator $S_{n-m,n-m}$ for which the contribution from the bridge cancels and the measured noise is proportional to the temperature average on either side: $S_{H,C} \propto \int_{H,C} dr\, T_e(r)$ (see Supplementary Section 4.1 for a detailed discussion). We achieve this by implementing differential noise thermometry[37], in which differential thermal noise is amplified, band-pass filtered, and frequency-integrated, resulting in a voltage signal proportional to the total noise power in a frequency band, with non-overlapping bands chosen for hot and cold sides (see Fig.1c, Methods section, and Supplementary Section 10). The geometry and circuit together allow for the isolation of the heat transport-induced nonlocal noise (see Supplementary Section 4.1). To achieve an accurate measurement of the transported energy, we must ensure low electron energy loss to phonons in the thermometers. Graphene possesses several properties that are well-suited to electronic thermometry[7,23,24,34], including strong interactions, exceptionally low energy loss, and small electronic thermal conductance (see Methods section). By defining local electrostatic gates for the hot and cold sides and exploiting tunable environmental disorder, graphene can be tuned to a diffusive regime, allowing for Joule heating and accurate noise thermometry on the hot and cold sides independent of the bridge state.

As a first demonstration of electronic thermal transport measurement using nonlocal noise, we employ graphene as a bridge connecting graphene thermometers in a monolithic multiterminal graphene device. Figure 1d shows an H-shaped graphene device encapsulated in insulating hexagonal boron nitride (hBN). The device is etched to define the hot, cold, and bridge regions. A low-frequency current is injected into the hot side, dissipating Joule power $P_H^J$. As the power increases, the measured noise power increases monotonically, resulting in a corresponding change in the measured temperature which is linear at low Joule power (Fig.1d, left panel). At two different electron densities of the bridge, the hot side temperature change is effectively identical, indicating that only a small fraction of the total applied Joule power is transported across the bridge. The temperature change on the cold side (Fig.1d, right panel), in contrast, is far smaller: for $\Delta T_H = 0.6$ K, we observe $\Delta T_C = 20\text{-}30$ mK, as expected for a small energy current across the bridge, and depends strongly on the bridge density.

Similar to the electrical conductance, the thermal conductance of the graphene bridge can be controlled by a voltage applied to a local gate. For this purpose, we fix the applied power in the linear response regime and tune the bridge electron density using a local metal gate on the bridge, $V_g^{bridge}$ (see Fig.2a, inset). The hot and cold side gates are held fixed at values that maintain the thermometers in a diffusive regime. The hot side temperature change (Fig.2a, top panel) is observed to be independent of the bridge density at three different bath temperatures. The cold side temperature change (Fig.2a, lower panel), in contrast, varies strongly as a function of the bridge gate, and shows a distinct trend that is reproduced at the three bath temperatures. Comparison with the four-point electrical resistance of the bridge (Fig.2b, upper panel) shows that the Dirac peak of the graphene bridge where resistance is maximal corresponds to the minimum in $\Delta T_C$, reflecting a thermal conductance modulation of the bridge with density.

To quantify the thermal conductance from the two measured temperatures, we require the energy current $Q_{bridge}$. Because energy loss to phonons in the graphene cold side thermometer is negligible, it can be shown that $Q_{bridge} = \frac{2}{3} G_C^{th} \Delta T_C$, where $G_C^{th}$ is the thermal conductance of the cold side graphene measured by self-heating[38] (see Supplementary Section 4.2 for the derivation). Crucially, the cold side graphene serves both as a thermometer and a power meter. This can be understood by considering the effective thermal circuit of the device (Fig.1b). In this model, $\Delta T_{H,C}$ can be computed as a function of the total input power $Q_{in}$ and the three thermal resistors, from which we obtain $G_{bridge}^{th} = G_{C'}^{th} \Delta T_{C'} / (\Delta T_{H'} - \Delta T_{C'})$, showing that $Q_{bridge} = G_{C'}^{th} \Delta T_{C'}$ (primed quantities refer to the circuit model; see Supplementary Section 5 for the connection between the thermal circuit and device). This result originates in the negligible energy loss to phonons. Thus, the temperature rise combined with the local thermal conductance accounts for all the power impinging on the cold side. This analysis can be extended to the case where electron-phonon coupling of the graphene thermometers is present, such as at high temperatures $T > 70$ K, since the electron-phonon energy loss can be directly measured in the same setup and accounted for quantitatively (see Supplementary Section 6)[7,23,24]. Here, we restrict ourselves to the low-temperature regime where electron-phonon coupling in the thermometers is negligible (see Supplementary Section 14 for an example at higher temperature where electron-phonon coupling is present).

The resulting thermal conductance of the bridge is shown in the middle panel of Fig.2b and exhibits strong anti-correlation with the electrical resistance. This observation can be made precise by computing a Lorenz ratio from the conductances, defined in analogy to the Wiedemann-Franz law as $\frac{L}{L_0} = \frac{G^{th}R}{T_{bath}}/L_0$, where $L_0 = \frac{\pi^2}{3}\left(\frac{k_B}{e}\right)^2$ is the Lorenz number. The Lorenz ratio at $T_{bath}=5$ K (Fig.2b, lower panel, blue curve) is close to 1 for the entire gate voltage range. This result demonstrates conclusively the thermal transport origin of the measured noise, validates the analysis methodology, and indicates the negligible effect of phonons, radiation and contact resistance in this regime (see Supplementary Sections 11 and 13).

At higher temperature, we find a density-dependent violation of the Wiedemann-Franz law indicating the breakdown of a simple diffusive electronic system. At $T_{bath}=20$ K and 30 K, the Lorenz ratio is suppressed away from the Dirac point, exhibiting a local minimum and saturating at an intermediate value. Electron-electron interactions are predicted to suppress the Lorenz ratio away from charge neutrality[39–45]. This is a sign of the onset of the hydrodynamic regime, recently discovered in graphene[46], in which electron-electron interactions scatter energy current while conserving charge current. Interactions combined with disorder lead to different signatures: with long-range disorder, the Lorenz ratio is suppressed at high density[41], while with short-range disorder it is suppressed in a lower density regime[42]. The local minimum and high-density suppression of the Lorenz ratio thus point to a disordered hydrodynamic regime[44,45].

We now turn to show that this method can be generalized to probe other low-dimensional materials. Graphene has previously been used as a contact intermediary for low-dimensional materials for which metallic contact is difficult to achieve[47–51]. By replacing the graphene bridge with a different material of interest, thermal contact may be established between the two graphene thermometers and a low-dimensional material bridge. To test this idea at its ultimate limit, we bridge two graphene thermometers with a carbon nanotube (NT), as shown in Fig.3a (we consider an insulating, bulk material in Supplementary Section 14). Carbon NTs are one-dimensional metals or semiconductors depending on their atomic structures[52]. We grow carbon NTs and individually characterize and incorporate them into graphene devices (see Methods Section)[53,54]. In these devices, the graphene thermometers are not covered by hBN in order to ensure electrical contact between the graphene and the NT. As a result, the graphene thermometers are more disordered than fully-encapsulated devices and experience more energy loss at our operating temperatures (see Supplementary Section 7). The thermal quantities presented here are thus lower bounds.

The two-point electrical and thermal conductance, $G_{NT}$ and $G_{NT}^{th}$, of a nanotube device (Device 1) are shown in Fig.3b as a function of voltage $V_g^{NT}$ applied to a local metal gate above the nanotube. At $T_{bath}$ = 70 K, the electrical conductance exhibits a global minimum at $V_g^{NT} \sim 15V$ corresponding to a small gap in the electronic spectrum. The thermal conductance exhibits a similar feature. As the temperature is lowered, rapid modulations are observed in both electrical and thermal conductance, becoming more pronounced at lower temperature. Throughout, we find that $G_{NT}$ and $G_{NT}^{th}$ closely follow each other. The rapidly varying oscillatory conductance is indicative of the onset of Coulomb blockade through the disordered, substrate-supported nanotube[55]. For temperatures above the Coulomb blockade regime, the electrical conductance is nearly temperature independent, evidencing the previously-discovered weak electron-phonon interactions in carbon NTs (see Supplementary Section 13). In a second device (Device 2) with a larger bandgap and higher channel resistance (Fig.3c), the device is in a disordered Coulomb blockade regime and exhibits sharper peaks alternating with vanishing electrical conductance at lower temperatures (see Fig.3c, bottom panel inset). The thermal conductance data trends with the electrical signal, despite the much higher channel resistance greater than 1 MΩ, corresponding to $G_{NT} \sim 10^{-3} \frac{e^2}{h}$. The corresponding thermal measurement operates down to ~1% of the thermal conductance quantum $\frac{\pi^2}{3} \frac{k_B^2}{h} T$ at 5 K (see Supplementary Section 12 for Device 1 data at 5 K). The measurement of electronic thermal transport in a system with far less than a single open quantum channel demonstrates the exceptional sensitivity of graphene noise thermometers in our experiment.

The relation between $G_{NT}$ and $G_{NT}^{th}$ can be described quantitatively by considering the Lorenz ratio $L_{NT}/L_0 = G_{NT}^{th}/L_0 T G_{NT}$. For Device 1, with higher conductance, the Lorenz ratio is significantly above 1 for all gate and temperature ranges measured (see lower panels of Fig.3b), indicating a Wiedemann-Franz violation. This is consistent with previous measurements in quasi-1D systems and with several theoretical predictions[3,56–58]. Close inspection shows that $L_{NT}/L_0$ exhibits peaks whenever the electrical conductance shows a dip, suggesting enhanced thermal conduction when electrical conduction is suppressed. In Device 2, with higher resistance, this enhanced thermal conduction is more clearly visible by plotting the inverse Lorenz ratio, $(L_{NT}/L_0)^{-1}$, which is strongly positively correlated with the electrical conductance (Fig. 3c). This gate dependence rules out an extrinsic contact resistance as the source of the Lorenz ratio behavior (see Supplementary Section 11). The observed correlation of $(L_{NT}/L_0)^{-1}$ and $G_{NT}$ suggests the presence of a channel of excess thermal conduction that does not rely on direct electron transport.

To account for a heat transport channel that is active even with suppressed electrical conduction, we propose a model for plasmon hopping mediated by Coulomb interactions. The long-range Coulomb interaction has measurable effect on many NT properties[12,59–61]. We consider a minimal model of a 1D conductor separated into two parts by an impenetrable barrier. In the absence of electron transport, energy transport by hot electrons cannot be achieved. However, a long-range interaction allows for energy transfer across the barrier even in the absence of direct charge tunneling. In this case, hot plasmons (density fluctuations) induce fluctuations across the electron barrier, leading to an energy current (Fig.4b, inset). With Coulomb interactions, the plasmon hopping energy current obeys $Q \propto T_H^2 - T_C^2$, while in the presence of screening by a nearby metal gate, the result is modified to $Q \propto T_H^4 - T_C^4$ (see Supplementary Section 8 for a detailed calculation).

We further test long-range plasmonic energy transport in the NT devices using nonlinear thermal transport. We relax the condition $\Delta T_H \lesssim T_{bath}$ of the previous measurement by measuring the nanotube energy current $Q_{NT}$ up to large thermal bias $\Delta T_H$. This measurement is the thermal analogue of a current-voltage curve in electrical measurement. Figure 4a shows $Q_{NT}$ as a function of $\Delta T_H/T_{bath}$ at representative gate voltages for Devices 1 and 2. We observe a superlinear increase of $Q_{NT}$ for all measured gate voltages and bath temperatures. Figure 4b shows a log-log plot of $Q_{NT} + Q_0$ versus $(\Delta T_H + T_{bath})/T_{bath}$, where $Q_0 = aT_{bath}^p$ is a fitting parameter with $a$ corresponding to the proportionality constant of the expression $Q_{NT} \propto T_H^p - T_C^p$. The highly linear scaling observed suggests that $Q_{NT}$ follows the power law behavior above with well-defined $p$. The slope of this plot provides $p$, which ranges between 2-6 depending on the nanotube resistance $R_{NT}$ (Fig. 4c). For the more resistive nanotube of Device 2, $R_{NT} \gg \frac{h}{e^2}$ (vertical dashed grey line), we observe $p \sim 4$, suggesting that once the electron transmission coefficient is far less than 1 and tunneling becomes suppressed, plasmon hopping via screened Coulomb interactions may be an important contribution to heat transport. For the highly-conductive NT of Device 1, $R_{NT} \sim \frac{h}{e^2}$ and $p$ ranges from 2 to 4. In this regime, electron transport is non-negligible, necessitating further theoretical modelling. We also note that our experimental observations cannot be described by an existing theory[62] in a disordered Luttinger regime with short-range interactions only, which predicts $Q \propto (T_H - T_C)^{\frac{4}{3}}$ and gives $p < 2$. At a high conductance and high temperature point, we find $p \sim 6$, indicating the possible presence of additional energy transport mechanisms. The observed superlinear behavior, and the associated exponents, are not explained by extrinsic effects like contact resistance (see Supplementary Section 11; other parasitic contributions are ruled out in Supplementary Section 13). Our observations motivate future work to understand the interplay of long-range interactions and 1D electron and heat transport.

In conclusion, we have demonstrated the measurement of nonlocal voltage fluctuations induced by electronic thermal transport. By using graphene noise thermometers, we show high-sensitivity electronic heat transport experiments in 2D van der Waals, 1D nanotubes and 0D localized systems, where we observe interaction effects in energy transport. In Supplementary Section 14, we demonstrate thermal transport in a microscale bulk electrical insulator, exhibiting measurable signals of magnetic thermal transport and a crossover to the phonon-coupled regime. Our approach enables the study of electronic thermal transport in a wide variety of low dimensional systems that were previously out of reach.

**Acknowledgments**

We thank Bertrand Halperin, Andrew Lucas, Sankar Das Sarma, Connie Mousatov, Kin Chung Fong, and Jesse Crossno for helpful discussions, James MacArthur for assistance with electronics design and construction, and Marine Arino and Hugo Bartolomei for their assistance in the early stages of this work. This work is supported by ARO (W911NF-17-1-0574) for developing RF technology and characterization, and ONR (N00014-16-1-2921) for device fabrication and measurements. A.V.T. acknowledges support from the DoD through the NDSEG Fellowship Program. J.W. and P.K. acknowledge support from NSF (DMR-1922172) for data analysis. K.W. and T.T. acknowledge support from the Elemental Strategy Initiative conducted by the MEXT, Japan, Grant Number JPMXP0112101001, JSPS KAKENHI Grant Number JP20H00354 and the CREST (JPMJCR15F3), JST. Work by K.A.M. at Argonne National Laboratory was supported by the U.S. Department of Energy, Office of Science, Basic Energy Sciences, Materials Sciences and Engineering Division.

**Author Contributions**

J.W. and P.K. conceived the experiments. J.W. performed the experiments and analyzed the data. L.E.A. fabricated the nanotube-graphene devices. J.W., Y.J.S, and D.H.N. fabricated the graphene devices. M.R. fabricated the α-RuCl$_3$ devices. X.F. and D.G.N. synthesized the bulk α-RuCl$_3$ crystals. T.T. and K.W. synthesized the bulk hBN crystals. B.S. performed nonlocal noise calculations. K.A.M. developed the plasmon hopping theory. J.W., L.E.A., A.V.T., Z.Y., M.R., B.S., K.A.M., and P.K. discussed the results and interpretations. J.W. and P.K. wrote the manuscript in consultation with the other authors.

**Competing Interests**

The authors declare no competing financial interests.

**Correspondence and requests for materials** should be addressed to P.K. (e-mail: pkim@physics.harvard.edu).

**References**

1. Franz, R. & Wiedemann, G. Ueber die Wärme-Leitungsfähigkeit der Metalle. *Ann. Phys.* **165**, 497–531 (1853).

2. Pines, D. & Nozières, P. *The Theory of Quantum Liquids*. (W.A. Benjamin, 1966).

3. Wakeham, N. *et al.* Gross violation of the Wiedemann-Franz law in a quasi-one-dimensional conductor. *Nat. Commun.* **2**, 396 (2011).

4. Smith, R. P. *et al.* Marginal breakdown of the Fermi-liquid state on the border of metallic ferromagnetism. *Nature* **455**, 1220–1223 (2008).

5. Tanatar, M. A., Paglione, J., Petrovic, C. & Taillefer, L. Anisotropic violation of the Wiedemann-Franz law at a quantum critical point. *Science* **316**, 1320–2 (2007).

6. Hill, R. W., Proust, C., Taillefer, L., Fournier, P. & Greene, R. L. Breakdown of Fermi-liquid theory in a copper-oxide superconductor. *Nature* **414**, 711–715 (2001).

7. Crossno, J. *et al.* Observation of the Dirac fluid and the breakdown of the Wiedemann-Franz law in graphene. *Science* **351**, 1058–61 (2016).

8. Enter 2D quantum materials. *Nat. Mater.* **19**, 1255–1255 (2020).

9. Deshpande, V. V, Bockrath, M., Glazman, L. I. & Yacoby, A. Electron liquids and solids in one dimension. *Nature* **464**, 209–16 (2010).

10. Zhou, Y. *et al.* Signatures of bilayer Wigner crystals in a transition metal dichalcogenide heterostructure. *arXiv* (2020).

11. Smoleński, T. *et al.* Observation of Wigner crystal of electrons in a monolayer semiconductor. (2020).

12. Shapir, I. *et al.* Imaging the electronic Wigner crystal in one dimension. *Science (80-. ).* **364**, 870–875 (2019).

13. Andrei, E. Y. & MacDonald, A. H. Graphene bilayers with a twist. *Nat. Mater.* **19**, 1265–1275 (2020).

14. Mak, K. F., Shan, J. & Ralph, D. C. Probing and controlling magnetic states in 2D layered magnetic materials. *Nature Reviews Physics* **1**, 646–661 (2019).

15. Li, M. & Chen, G. Thermal transport for probing quantum materials. *MRS Bull.* **45**, 348–356 (2020).

16. Tritt, T. M. *Thermal Conductivity*. (Kluwer Academic Publishers-Plenum Publishers, 2004).

17. Banerjee, M. *et al.* Observed quantization of anyonic heat flow. *Nature* **545**, 75–79 (2017).

18. Banerjee, M. *et al.* Observation of half-integer thermal Hall conductance. *Nature* 1 (2018). doi:10.1038/s41586-018-0184-1

19. Srivastav, S. K. *et al.* Universal quantized thermal conductance in graphene. *Sci. Adv.* **5**, eaaw5798 (2019).

20. Dutta, B. *et al.* Thermal Conductance of a Single-Electron Transistor. *Phys. Rev. Lett.* **119**, 077701 (2017).

21. Cui, L. *et al.* Quantized thermal transport in single-atom junctions. *Science (80-. ).* **355**, 1192–1195 (2017).

22. Mosso, N. *et al.* Heat transport through atomic contacts. *Nat. Nanotechnol.* **12**, 430–433 (2017).

23. Crossno, J., Liu, X., Ohki, T. A., Kim, P. & Fong, K. C. Development of high frequency and wide bandwidth Johnson noise thermometry. *Appl. Phys. Lett.* **106**, 023121 (2015).

24. Fong, K. C. *et al.* Measurement of the Electronic Thermal Conductance Channels and Heat Capacity of Graphene at Low Temperature. *Phys. Rev. X* **3**, 041008 (2013).

25. Chiatti, O. *et al.* Quantum thermal conductance of electrons in a one-dimensional wire. *Phys. Rev. Lett.* **97**, (2006).

26. Molenkamp, L. W. *et al.* Peltier coefficient and thermal conductance of a quantum point contact. *Phys. Rev. Lett.* **68**, 3765–3768 (1992).

27. Seol, J. H. *et al.* Two-dimensional phonon transport in supported graphene. *Science (80-. ).* **328**, 213–216 (2010).

28. Kim, P., Shi, L., Majumdar, A. & McEuen, P. L. Thermal transport measurements of individual multiwalled nanotubes. *Phys. Rev. Lett.* **87**, 215502-1-215502–4 (2001).

29. Mosso, N. *et al.* Thermal Transport through Single-Molecule Junctions. *Nano Lett.* **19**, 7614–7622 (2019).

30. Cui, L. *et al.* Thermal conductance of single-molecule junctions. *Nature* **572**, 628–633 (2019).

31. Johnson, J. B. Thermal agitation of electricity in conductors. *Phys. Rev.* **32**, 97–109 (1928).

32. Nyquist, H. Thermal agitation of electric charge in conductors. *Phys. Rev.* **32**, 110–113 (1928).

33. Qu, J. F. *et al.* Johnson noise thermometry. *Meas. Sci. Technol* **30**, 112001 (2019).

34. Fong, K. C. & Schwab, K. C. Ultrasensitive and Wide-Bandwidth Thermal Measurements of Graphene at Low Temperatures. *Phys. Rev. X* **2**, 031006 (2012).

35. Yiğen, S. & Champagne, A. R. Wiedemann-franz relation and thermal-transistor effect in suspended graphene. *Nano Lett.* **14**, 289–293 (2014).

36. Sukhorukov, E. V. & Loss, D. Noise in multiterminal diffusive conductors: Universality, nonlocality, and exchange effects. *Phys. Rev. B* **59**, 13054–13066 (1999).

37. Talanov, A. V., Waissman, J., Taniguchi, T., Watanabe, K. & Kim, P. High-bandwidth, variable-resistance differential noise thermometry. *Rev. Sci. Instrum.* **92**, 014904 (2021).

38. Pozderac, C. & Skinner, B. Relation between Johnson Noise and heating power in a two-terminal conductor. Preprint available at http://arxiv.org/abs/2104.05714 (2021).

39. Principi, A. & Vignale, G. Violation of the Wiedemann-Franz Law in Hydrodynamic Electron Liquids. *Phys. Rev. Lett.* **115**, 056603 (2015).

40. Lucas, A. & Das Sarma, S. Electronic hydrodynamics and the breakdown of the Wiedemann-Franz and Mott laws in interacting metals. *Phys. Rev. B* **97**, 245128 (2018).

41. Li, S., Levchenko, A. & Andreev, A. V. Hydrodynamic electron transport near charge neutrality. *Phys. Rev. B* **102**, 075305 (2020).

42. Xie, H.-Y. & Foster, M. S. Transport coefficients of graphene: Interplay of impurity scattering, Coulomb interaction, and optical phonons. *Phys. Rev. B* **93**, 195103 (2016).

43. Lucas, A., Davison, R. A. & Sachdev, S. Hydrodynamic theory of thermoelectric transport and negative magnetoresistance in Weyl semimetals. *Proc. Natl. Acad. Sci. U. S. A.* **113**, 9463–9468 (2016).

44. Zarenia, M., Principi, A. & Vignale, G. Disorder-enabled hydrodynamics of charge and heat transport in monolayer graphene. *2D Mater.* **6**, 035024 (2019).

45. Zarenia, M., Smith, T. B., Principi, A. & Vignale, G. Breakdown of the Wiedemann-Franz law in AB-stacked bilayer graphene. *Phys. Rev. B* **99**, 161407 (2019).

46. Lucas, A. & Fong, K. C. Hydrodynamics of electrons in graphene. *Journal of Physics Condensed Matter* **30**, 53001 (2018).

47. Cui, X. *et al.* Multi-terminal transport measurements of MoS2 using a van der Waals heterostructure device platform. *Nat. Nanotechnol.* **10**, 534–540 (2015).

48. Liu, Y. *et al.* Toward barrier free contact to molybdenum disulfide using graphene electrodes. *Nano Lett.* **15**, 3030–3034 (2015).

49. Wang, K. *et al.* Electrical control of charged carriers and excitons in atomically thin materials. *Nat. Nanotechnol.* **13**, 128–132 (2018).

50. El Abbassi, M. *et al.* Robust graphene-based molecular devices. *Nature Nanotechnology* **14**, 957–961 (2019).

51. Yang, C., Qin, A., Tang, B. Z. & Guo, X. Fabrication and functions of graphene-molecule-graphene single-molecule junctions. *J. Chem. Phys.* **152**, 120902 (2020).

52. Ilani, S. & McEuen, P. L. Electron Transport in Carbon Nanotubes. *Annu. Rev. Condens. Matter Phys.* **1**, 1–25 (2010).

53. Sfeir, M. Y. *et al.* Optical spectroscopy of individual single-walled carbon nanotubes of defined chiral structure. *Science (80-. ).* **312**, 554–556 (2006).

54. Cheng, A., Taniguchi, T., Watanabe, K., Kim, P. & Pillet, J. D. Guiding Dirac Fermions in Graphene with a Carbon Nanotube. *Phys. Rev. Lett.* **123**, 216804 (2019).

55. McEuen, P., Bockrath, M., Cobden, D., Yoon, Y.-G. & Louie, S. Disorder, Pseudospins, and Backscattering in Carbon Nanotubes. *Phys. Rev. Lett.* **83**, 5098–5101 (1999).

56. Garg, A., Rasch, D., Shimshoni, E. & Rosch, A. Large Violation of the Wiedemann-Franz Law in Luttinger Liquids. *Phys. Rev. Lett.* **103**, 096402 (2009).

57. Kane, C. L. & Fisher, M. P. A. Thermal Transport in a Luttinger Liquid. *Phys. Rev. Lett.* **76**, 3192–3195 (1996).

58. Li, M.-R. & Orignac, E. Heat conduction and Wiedemann-Franz law in disordered Luttinger liquids. *Europhys. Lett.* **60**, 432–438 (2002).

59. Deshpande, V. V. & Bockrath, M. The one-dimensional Wigner crystal in carbon nanotubes. *Nat. Phys.* **4**, 314–318 (2008).

60. Pecker, S. *et al.* Observation and spectroscopy of a two-electron Wigner molecule in an ultraclean carbon nanotube. *Nat. Phys.* **9**, 1–17 (2013).

61. Shi, Z. *et al.* Observation of a Luttinger-liquid plasmon in metallic single-walled carbon nanotubes. *Nat. Photonics* **9**, 515–519 (2015).

62. Fazio, R., Hekking, F. W. J. & Khmelnitskii, D. E. Anomalous Thermal Transport in Quantum Wires. *Phys. Rev. Lett.* **80**, 5611–5614 (1998).

**Methods**

*Nanotube-Graphene Device Fabrication*

Carbon NTs are grown in a CVD furnace using the methods described in Ref.[1]. The growth substrate is a 5x5 mm$^2$ silicon chip with a slit in the center, oriented perpendicular to the gas flow direction. Catalyst is applied on one side of the slit, so that NTs grow suspended across the slit. Suspended NTs were found and characterized using Rayleigh scattering spectroscopy and imaging[2]. By matching peaks in Rayleigh scattering intensity with NT optical transition energies, the chiral indices (and thus diameter and metallic/semiconducting nature) of the NT can be determined.

Heterostructures of monolayer graphene on top of a 20-60 nm BN flake were prepared using the inverted stacking technique. A 200-500 nm wide, $> 10\ \mu m$ long slit was then created in the graphene by defining a polymethyl methacrylate (PMMA) mask with electron beam (e-beam) lithography and etching with O$_2$ plasma in a reactive ion etcher. A second e-beam lithography step defines a resist-free window above the heterostructure, while the rest of the chip remains coated in $\sim 100\ nm$ of resist. The growth chip and PMMA-coated sample are pressed together until mechanical contact is observed, then heated to 180 $°C$ for 5 minutes to melt the resist. The chips are then cooled to 90 $°C$ and slowly separated. Successful NT transfer is confirmed by electron microscope or atomic force microscope imaging.

Following NT transfer, electrical contacts were made to the edges of the graphene following the method in Ref.[3]. The unwanted sections of the heterostructure are removed by reactive ion etching with CHF$_3$. An insulating layer of 120 nm SiO$_2$ was made above the NT by e-beam lithography of hydrogen silsesquioxane (HSQ) resist and development with CD-26 developer. A final e-beam lithography step defined the mask for the local top gate above the NT, which was formed by thermal evaporation of 3 nm Cr/7 nm Pd/70 nm Au (using an angled, rotating stage to mitigate height differences between different parts of the structure).

*Differential Noise Thermometry with Graphene*

Each thermometer should measure the local temperature $T_{H,C}$ without cross-contamination of signals. Single-ended amplification of $S_{H,C}$ would mix signals from either side due to a common ground and cause direct Joule heating of the cold side if the bridge is electrically conducting (see Supplementary Section 3). We therefore implement a differential noise thermometry measurement, described in detail elsewhere[4]. Briefly, each terminal pair is connected to a balanced matching circuit that couples high frequency signals into a differential low noise amplifier (see Fig.1c). The resonant frequencies for the two matching circuits, between 100 MHz-1 GHz, are chosen to be separated in frequency by several times the circuit bandwidth, so that the hot and cold noise signals are mutually filtered and cross-correlations are suppressed. The amplified signals are bandpass filtered and amplified at a second stage and sent through a power detector which generates a voltage proportional to the integrated high-frequency noise spectral density (see Supplementary Section 10). By applying a low-frequency current at frequency *f*, the system is heated by Joule power at frequency *2f*, and the output voltage is amplitude-modulated at frequency *2f*. Using lock-in amplifiers, we isolate the change in noise power amplitude due to the applied Joule power. After calibration (see Supplementary Section 1), the *2f* noise power voltage signal is converted into a temperature rise, $\Delta T_{H,C}$. Previously, we have shown this measurement can achieve sub-milliKelvin precision in 30s averaging time[4].

For thermal conductance measurement, the noise thermometer should have negligible energy loss. Graphene possesses a combination of properties that suit this purpose. Strong electron-electron interactions allow for thermalized temperature distributions down to sub-micrometer length scales[5,6]. The light carbon lattice and stiff bonding result in weak electron-optical phonon coupling, while the large mismatch between the Fermi and sound velocities puts acoustic phonons in the quasi-elastic scattering regime in which energy loss is low[7]. The small Fermi surface around the Dirac point yields negligible umklapp scattering, and the exceptional chemical cleanliness means inelastic impurity scattering is largely absent. Because of its 2D nature, it possesses small electronic thermal conductance compared to 3-dimensional bulk materials and is thus sensitive to small quantities of injected energy[8–10].

**Methods References**

1. Sfeir, M. Y. *et al.* Optical spectroscopy of individual single-walled carbon nanotubes of defined chiral structure. *Science (80-. ).* **312**, 554–556 (2006).

2. Cheng, A., Taniguchi, T., Watanabe, K., Kim, P. & Pillet, J. D. Guiding Dirac Fermions in Graphene with a Carbon Nanotube. *Phys. Rev. Lett.* **123**, 216804 (2019).

3. Wang, L. *et al.* One-dimensional electrical contact to a two-dimensional material. *Science (80-. ).* **342**, 614–617 (2013).

4. Talanov, A. V., Waissman, J., Taniguchi, T., Watanabe, K. & Kim, P. High-bandwidth, variable-resistance differential noise thermometry. *Rev. Sci. Instrum.* **92**, 014904 (2021).

5. Müller, M., Schmalian, J. & Fritz, L. Graphene: A Nearly Perfect Fluid. *Phys. Rev. Lett.* **103**, 025301 (2009).

6. Tielrooij, K. J. *et al.* Photoexcitation cascade and multiple hot-carrier generation in graphene. *Nat. Phys.* **9**, 248–252 (2013).

7. Crossno, J. *et al.* Observation of the Dirac fluid and the breakdown of the Wiedemann-Franz law in graphene. *Science* **351**, 1058–61 (2016).

8. Walsh, E. D. *et al.* Graphene-based josephson-junction single-photon detector. *Phys. Rev. Appl.* **8**, 024022 (2017).

9. Lee, G.-H. *et al.* Graphene-based Josephson junction microwave bolometer. *Nature* **586**, 42–46 (2020).

10. Kokkoniemi, R. *et al.* Bolometer operating at the threshold for circuit quantum electrodynamics. *Nature* **586**, 47–51 (2020).

**Figure 1**: **Nonlocal Noise Thermometry in Multiterminal Devices. a**, Schematic of multiterminal noise measurement. A diffusive, conducting electron system is connected to terminals held at a bath temperature $T_{bath}$. Current is injected into one terminal and escapes via a ground, while other terminals are floating. Finite-element simulation of the temperature and current distribution assuming a uniform conductivity is shown in the color scale and streamlines, respectively. In the absence of energy loss to phonons, the noise $S_{nm}$ measured at any two terminals is given by a weighted function of the electronic temperature distribution $T_e(\mathbf{r})$, where $\mathbf{r}$ is a location in the conductor (see main text). **b**, Thermal circuit for a thermal conductance measurement. Joule power $P_H^J$ is injected into a hot side reservoir connected by thermal resistance $R_H^{th}$ to $T_{bath}$. The bridge thermal resistance $R_{bridge}^{th}$ allows thermal current $Q$ to cross from hot to cold side, connected to the bath by $R_C^{th}$. **c**, Circuit and geometry for nonlocal noise thermometry. Box I (blue, center): finite element simulation of the device geometry. Joule power is dissipated on the hot side (left, wide rectangle) due to injected current density (streamlines). Thermal current is transported across the bridge while the electrical current across the bridge $I_{bridge} = 0$, causing heating of the cold side (right, narrow rectangle). Box II (green), box III (yellow), box IV (red), correspond respectively to the balanced matching circuit, balanced current excitation and resistance measurement, and noise measurement amplification chain (see Methods section for details). **d**, The upper inset of center panel shows an optical image of the device (shown before top gate deposition for clarity), scale bar corresponds to 1μm. Schematic measurement setup overlaid on optical image shows the hot side current excitation and hot/cold side noise measurement simplified from panel c. The device stack consists of graphene encapsulated in hBN layers (lower inset). Left (right) panel displays hot (cold) side noise power and calibrated temperature change $\Delta T_H$ ($\Delta T_C$) versus applied Joule power, $P_H^J$, at $T_{bath} = 4.8K$ and fixed hot and cold side gate voltages at two values of the bridge density (solid and dashed lines). On the hot side (left panel), $\Delta T_H$ rises nearly identically, owing to a small amount of power escaping across the bridge, while $\Delta T_C$, which is far smaller, strongly depends on the bridge gate voltage, evidencing a difference in thermal current across the bridge.

**Figure 2: Electronic Thermal Conductance of Graphene. a**, Top panel: Hot side temperature change $\Delta T_H$ versus bridge gate, $V_g^{bridge}$, at fixed hot and cold side gate voltages $V_g^{hot}$ and $V_g^{cold}$. $T_{bath} = 5, 20,$ and $30K$ for blue, yellow, and red curves, respectively (for all panels in this figure). Lower panel: Corresponding cold side temperature change $\Delta T_C$ versus bridge gate, $V_g^{bridge}$. Inset: optical image of the hBN-encapsulated graphene device after topgate deposition. Scale bar: 1µm. Schematic circuit diagram: current is injected to the hot side at frequency $f$, and the resulting modulated noise power on hot and cold sides are measured yielding the temperature changes $\Delta T_H^{2f}$ and $\Delta T_C^{2f}$. **b**, Top panel display: Four-point electrical resistance of the bridge, $R_{bridge}$, versus $V_g^{bridge}$ (bottom axis) and bridge carrier density $n_{bridge}$ (top axis). In this temperature range, the resistance is nearly temperature independent. An excess resistance near $V_g^{bridge}=0$ at the lowest measured temperature arises due to induced disorder of the atomic layer deposited (ALD) insulating layer for the top gates. Middle panel shows thermal conductance of the bridge, $G_{bridge}^{th}$, deduced from the temperature changes and the independently measured cold side thermal conductance (see main text). Bottom panel shows Lorenz ratio of the bridge, $L_{bridge}/L_0$ (see main text for definition and discussion).

**Figure 3: Electronic Thermal Conductance of Carbon Nanotubes. a**, Left panel: Device schematic. A bottom hBN layer supports monolayer graphene, with an etched gap. A carbon NT connects the two graphene patches, which form the two noise thermometers. A local metallic topgate tunes the NT carrier density. Right panel: composite optical and scanning electron microscope image of the device. Scale bar shows 1µm. Dashed yellow line shows location of metal top gate. **b**, Two-point electrical and thermal conductance of a small-bandgap NT in Device 1. Main plots: electrical conductance in blue, thermal conductance in orange. Lower plots: Lorenz ratio $L_{NT}/L_0$ in dark green (see main text for definition). Electrical conductance shown in upper panel is replotted in faint blue dotted line for explicit comparison with $L_{NT}/L_0$. All thermal quantities are lower bounds (see main text and Supplementary Section 7). Top, middle, and bottom panels correspond to $T_{bath} = 70, 40$, and 20K respectively. Thermal bias for all plots $\Delta T_H/T_{bath} = 0.1$. See Supplementary Section 12 for data at $T_{bath} = 5$K. **c**, Two-point electrical and thermal conductance measured in Device 2 with a high-resistance NT bridge. Main plots: electrical conductance in blue, thermal conductance in orange. Lower plots: Inverse Lorenz ratio, $(L_{NT}/L_0)^{-1}$, in light green (see main text for definition), electrical conductance is reproduced in faint blue dotted line. Top, middle, and bottoms panels correspond to $T_{bath} = 15, 10$, and 5K respectively. Top and middle panels: $\Delta T_H/T_{bath} = 1$. Bottom panel: $\Delta T_H/T_{bath} = 0.5$. $30 mV$ DC voltage is applied across the NT to overcome the contact barrier; the corresponding measured DC current on the $\sim nA$ level leads to negligible heating at DC and does not affect the measurements at the *2f* heating frequency. All NT thermal quantities are lower bounds (see main text and Supplementary Section 7). Insets show zoom of Coulomb peaks. See Supplementary Section 12 for wider gate ranges.

**Figure 4: Nonlinear Thermal Transport in Carbon Nanotubes. a**, Thermal current across the NT $Q_{NT}$ versus scaled thermal bias, $\Delta T_H/T_{bath}$. Device 1: Orange: $T_{bath} = 70K$, $V_g^{NT} = 0V$. Yellow: $T_{bath} = 40K$, $V_g^{NT} = -10V$. Device 2: Purple: $T_{bath} = 50K$, $V_g^{NT} = -19.4V$ (multiplied by 20 for comparison). Red lines: fit to plasmon hopping model (see main text). **b**, Log-log linearizing plot of NT thermal current versus thermal bias. Device 1: Orange, yellow, and purple are identical to **a**. Blue and green data sets correspond to Device 1, $T_{bath} = 6K$, $V_g^{NT} = -10V$ and Device 2 $T_{bath} = 30K$, $V_g^{NT} = -19.9V$, respectively. Brown lines: fit to plasmon hopping model. Inset shows a schematic diagram of the plasmon hopping process, in which thermal electron density fluctuations are coupled across a barrier by long-range interactions, allowing the transport of energy even in an insulating system. **c**, Extracted exponents of the plasmon hopping model fit versus NT resistance $R_{NT}$. Symbols x and o correspond to data from Device 1 and Device 2, respectively. For device 1, blue, purple and red correspond to $T_{bath}=$ 6, 40, and 70K, respectively. For device 2, blue and purple correspond to $T_{bath}=$ 30 and 50K, respectively. Symbols without error bars have statistical error less than the symbol size. Dashed vertical grey line corresponds to $R_{NT} = h/e^2$.

# Figure 1

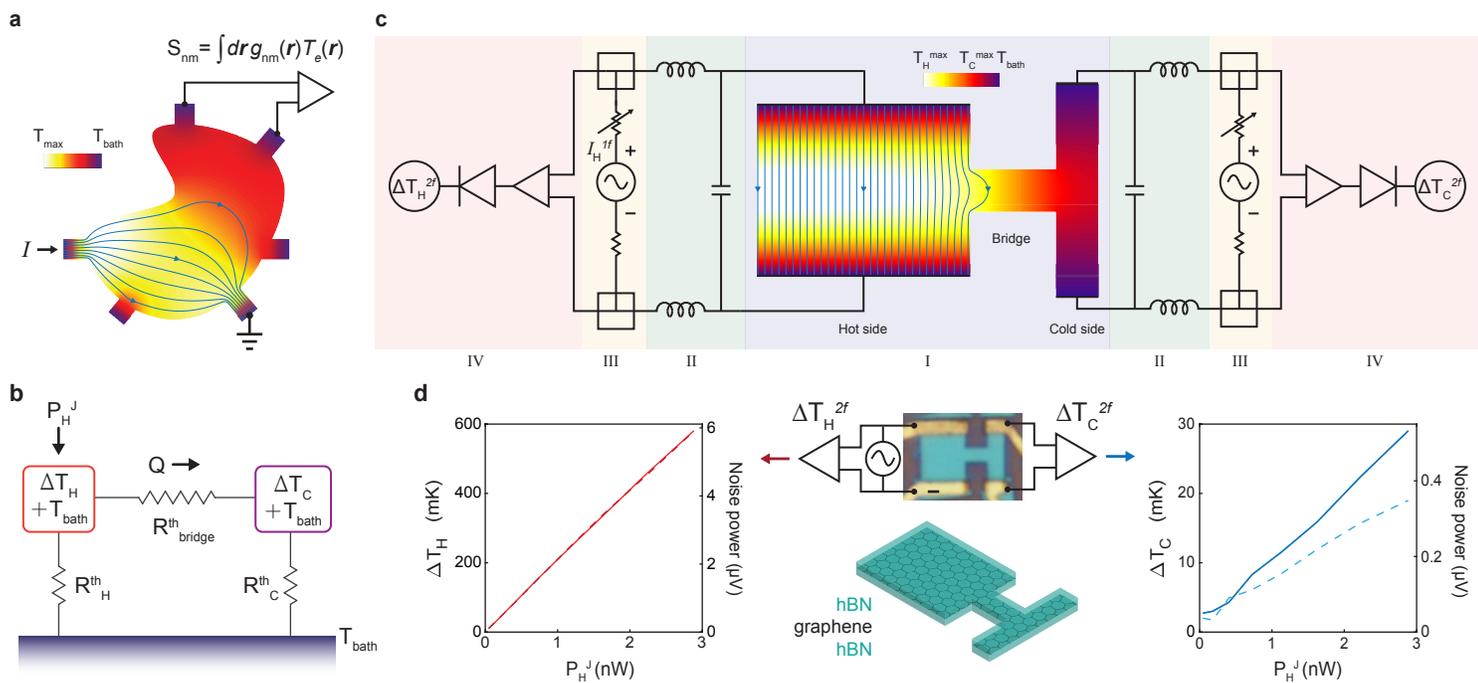

# Figure 2

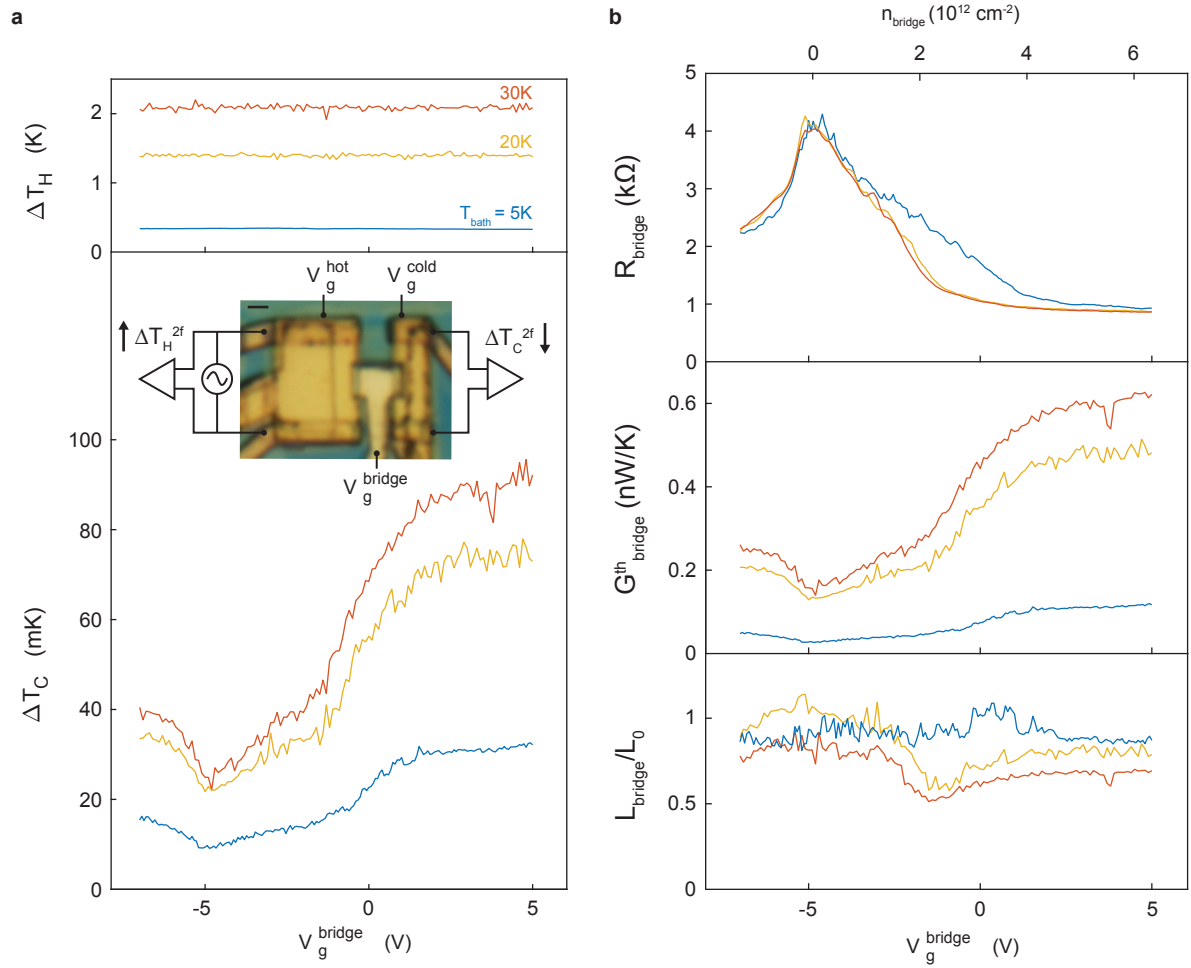

# Figure 3

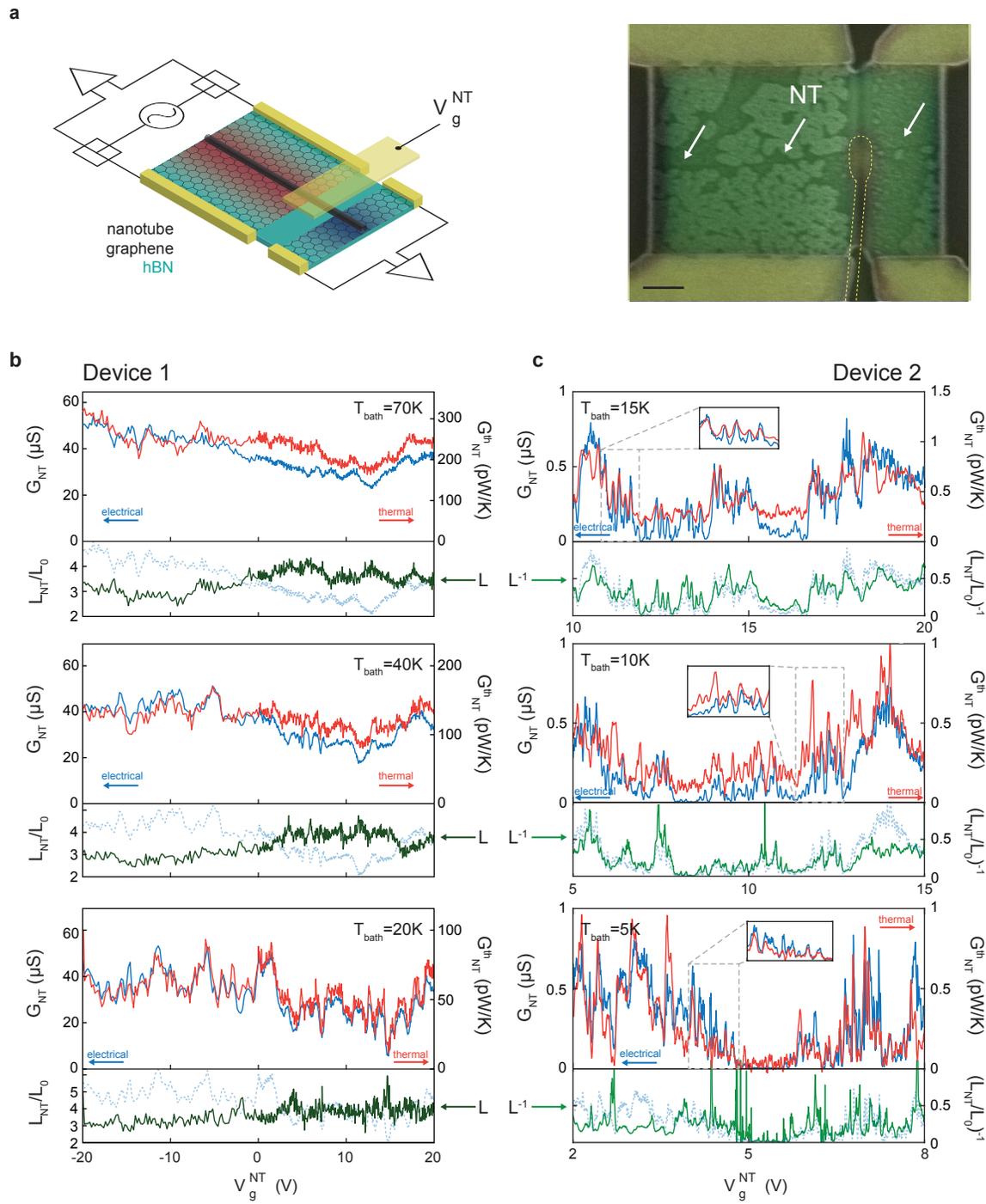

# Figure 4

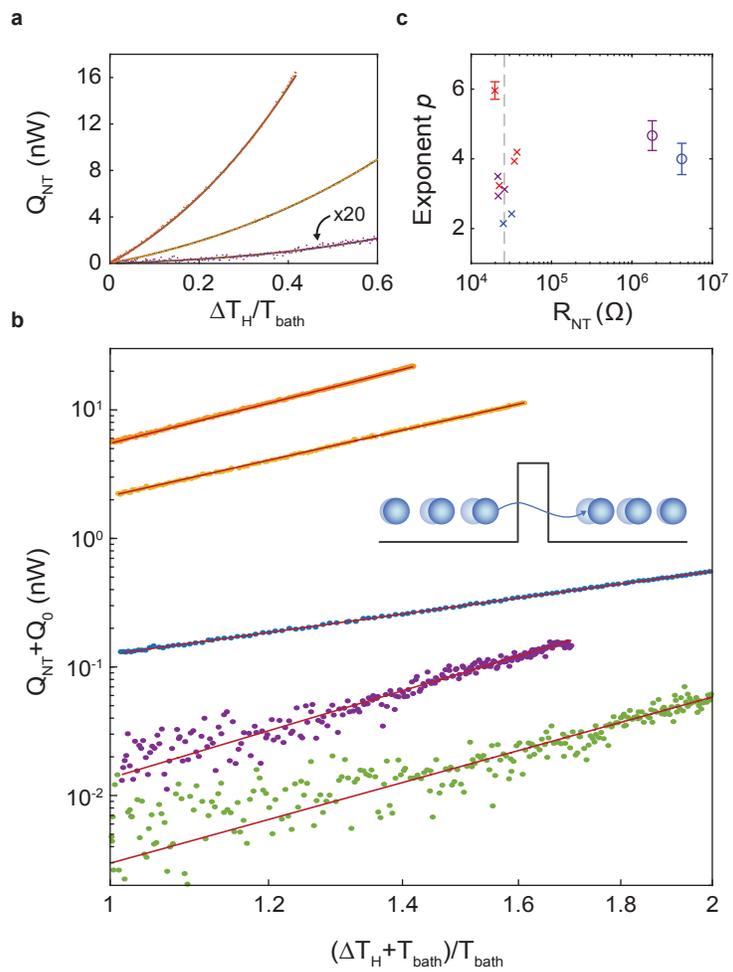

# Supplementary Information

for

## Electronic Thermal Transport Measurement in Low-Dimensional Materials with Graphene Nonlocal Noise Thermometry

Jonah Waissman[1], Laurel E. Anderson[1], Artem V. Talanov[1,2], Zhongying Yan[1], Young J. Shin[1], Danial H. Najafabadi[1], Mehdi Rezaee[2], Xiaowen Feng[3], Daniel G. Nocera[3], Takashi Taniguchi[4], Kenji Watanabe[5], Brian Skinner[6], Konstantin A. Matveev[7], Philip Kim[1,2]*

[1]*Department of Physics, Harvard University, Cambridge, MA 02138, USA*
[2]*John A. Paulson School of Engineering and Applied Sciences, Harvard University, Cambridge, MA 02138, USA*
[3]*Department of Chemistry and Chemical Biology, Harvard University, Cambridge, MA 02138 USA*
[4]*International Center for Materials Nanoarchitectonics, National Institute for Materials Science, 1-1 Namiki, Tsukuba 305-0044, Japan*
[5]*Research Center for Functional Materials, National Institute for Materials Science, 1-1 Namiki, Tsukuba 305-0044, Japan*
[6]*Department of Physics, The Ohio State University, Columbus, OH 43210, USA*
[7]*Materials Science Division, Argonne National Laboratory, Argonne, IL 60439, USA*

## Contents

# 1. Temperature Calibration

In this section, we describe the calibration that converts noise power to temperature. We use the calibration scheme described in Ref.[1] which we summarize here.

The integrated noise power measured by the power detector (see Fig1 of the main text) can be expressed as

$$P = G(T_e^{gr} + T_N) \tag{1.1}$$

where $P$ is the power detector output in Volts, $G$ is an effective gain constant, $T_e^{gr}$ is the electron temperature of the graphene thermometer being measured, and $T_N$ is an effective noise temperature. The quantities $G$ and $T_N$ are proportional to frequency integrals of the frequency-dependent gain and noise temperature over the system bandwidth. Both $G$ and $T_N$ depend on the resistance $R$ of the thermometer, such that $G = G(R)$ and $T_N = T_N(R)$. To accurately translate between noise power output and temperature, these two quantities must be known.

To obtain the gain and noise temperature, we perform calibration sweeps of the graphene thermometers. These involve sweeping a gate voltage to drive the thermometer across a range of resistance variation and repeating this sweep at several fixed temperatures. An example is shown in Fig.S1a, for a graphene thermometer used in Fig.3. As the graphene resistance changes due to the backgate voltage $V_{bg}$, the output noise power modulates, with the specific shape depending on the matching circuit design and sample resistance. With these curves, we plot noise power versus temperature at a fixed resistance, shown in Fig.S1b. The linear behavior is shown for three different values of the graphene resistance and is linear throughout the measured temperature range. We fit the expression above to the data, extracting $G$ from the slope and and $T_N$ from the x-axis intercept. The results are shown in Fig.S1c. The gain exhibits a local maximum, while the noise temperature shows a corresponding local minimum. These extremal points correspond to the best-matched resistance value in the circuit, where the graphene resistance is closest to the ideal match point of the matching circuit. From these data, noise power can be accurately converted to temperature.

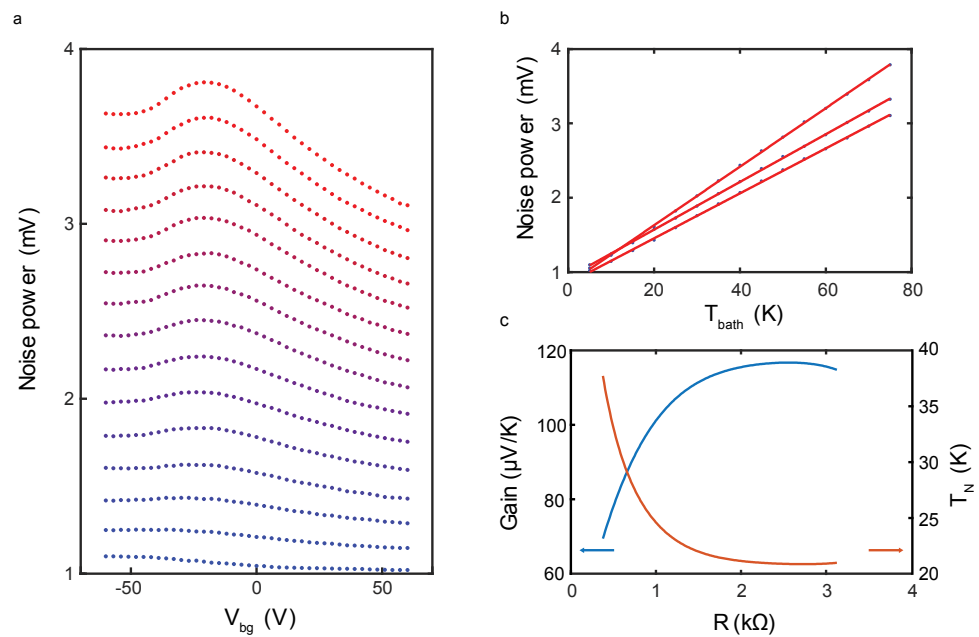

**Figure S1 Temperature Calibration.** a) Total noise power output of a graphene noise thermometer versus $V_{bg}$, for different temperatures $T_{bath}$ from $5\ K$ (blue) to $75\ K$ (red), spaced by 5 K. b) Total noise power versus $T_{bath}$ at fixed device resistance, shown for three different resistance values. Blue points: data, red lines: linear fit to Eq.1.1. c) Extracted gain and noise temperature $T_N$ versus device resistance $R$. Blue: gain (left axis), orange: $T_N$ (right axis).

## 2. Balancing Circuit

The thermal transport device described in the main text allows for independent thermal and electrical biasing of the bridge. In this section we briefly describe how we achieve the balanced circuit condition that allows for this flexible operation.

Thermal conductance is defined in the absence of electrical current flow. During a thermal transport measurement, the current injected to the hot side deposits Joule power, raising the electronic temperature and accomplishing a thermal bias for the bridge. The voltage bias of the bridge depends on the voltage distribution in the hot and cold sides. These in turn depend on the resistance to ground of all four contacts. These resistances will thus determine the potential distribution and the total electrical bias of the bridge.

Therefore, to establish a thermal bias without charge current, we balance the biasing circuit prior to the thermal measurements. A simplified schematic of the tuning circuit is shown in Fig.S2. A differential lock-in excitation is applied to the hot side. The cold side terminals are shorted together and their common voltage level $V_{unbal}$ is measured at the same lock-in frequency as the hot side excitation. The tunable biasing resistor is then adjusted to zero $V_{unbal}$. This scheme allows for thermal biasing without electrical current, but it is also more general. It allows for the application of arbitrary thermal and electrical bias, which can be independently tuned for a wider variety of experimental conditions.

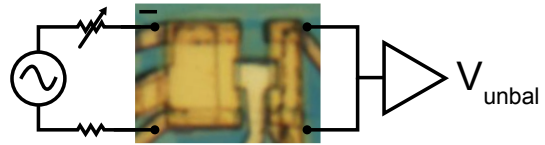

**Figure S2 Balancing circuit.** A balanced differential voltage excitation is applied to the hot side graphene thermometer via two biasing resistors. The balanced condition is attained by tuning one of the resistors. The opposite cold side graphene thermometer has its two contacts shorted together and measured as the unbalance voltage $V_{unbal}$. The tunable resistor is then adjusted until $V_{unbal} = 0$. Central image: optical image of the device after topgate deposition. Scale bar $1\mu m$.

## 3. Single-ended vs. Differential Amplification

Here, we briefly discuss the problems with single-ended amplification and the need for a differential approach. We seek to realize the effective thermal circuit shown in Fig.1b of the main text, reproduced in Fig.S3a. Naively, we can construct an electrical circuit for noise thermometry that resembles the thermal circuit using amplifiers referenced to a common ground. However, we will show below that this single-ended approach poses important issues.

Consider the electrical circuit shown in Fig.S3b, which represents an experimental scheme for noise measurement with single-ended amplification. Here, the resistors form a voltage divider for the voltage applied to the hot side, $V_{in}$.

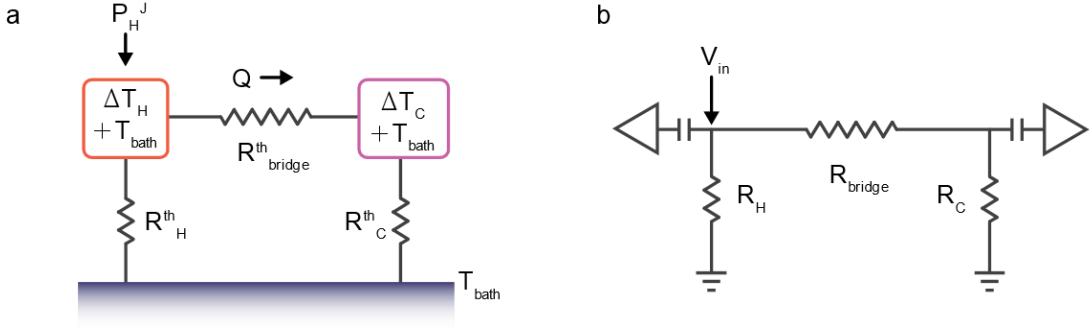

**Figure S3 Single-ended amplification. a)** Thermal circuit for a thermal conductance measurement, reproduced from Fig.1b of the main text. **b)** Electrical circuit under single-ended amplification. The hot side excitation $V_{in}$ is applied to the hot side resistor $R_H$, and flows to the hot side and cold side grounds. The resulting Joule heat is applied to all resistive elements of the circuit. The single-ended amplifiers measure voltage to ground, which includes fluctuations from all elements of the circuit. Thus, neither the Joule heat nor the noise thermometry is isolated for the hot and cold sides.

Two problems present themselves in this setup. First, there are two current paths to ground. The cold side is thus heated by direct Joule heating from electrical current flowing across the bridge, in addition to energy current due to the thermal conductance of the bridge. Second, the amplifiers measure the fluctuations associated with two current paths to ground, and thus do not isolate the desired $\Delta T_C$ and $\Delta T_H$, giving instead a mixed signal that is difficult to disentangle.

The differential circuit shown in Fig.1c of the main text plays two roles. First, it allows for thermally biasing the bridge without directly Joule heating the bridge and the cold side. Second, it isolates the voltage fluctuations such that the amplifiers measure the temperature rise associated with their respective thermometers exclusively, without mixing of signals from both sides.

# 4. Nonlocal Thermometry with Multiterminal Noise: Theory

In this section, we present theoretical calculations of multiterminal noise that undergird its use for nonlocal thermometry and thermal transport measurement.

## 4.1 Connecting Multiterminal Noise and Noise Thermometry

Here, we briefly review the formalism of Sukhorukov and Loss[2] for the hot electron regime and show how multiterminal noise can be related to the average of the thermometer temperature distribution by judicious choice of geometry and use of differential noise measurement.

Consider a multiterminal geometry like that shown in Fig.1a of the main text. The noise current on each terminal, $\delta I_n(t)$, gives rise to a correlation between the two terminals defined as $S_{nm} = \int_{-\infty}^{\infty} dt \langle \delta I_n(t) \delta I_m(t) \rangle$, which corresponds to the mean square noise current per unit frequency. Sukhorukov and Loss[2] showed that this noise current can be rewritten as

$$S_{nm} = \int d\boldsymbol{r} \nabla \phi_n \cdot \hat{\sigma} \nabla \phi_m \, \Pi(\boldsymbol{r}) \qquad (4.1.1)$$

Here, $\hat{\sigma}$ is the local conductivity tensor, $\phi_n$ is a characteristic potential associated with each terminal of the device, defined to obey a sum rule $\sum_n \phi_n(r) = 1$ such that the potential throughout the device $V(r)$ is given by $V(r) = \sum_n \phi_n(r) V_n$ where $V_n$ is the boundary potential of the $n^{th}$ terminal, and $\Pi(r)$ is a local effective electron temperature defined by $\Pi(r) = 2\int d\varepsilon f_0(\varepsilon, r)[1 - f_0(\varepsilon, r)]$ where $f_0(\varepsilon, r)$ is the energy and space dependent symmetric part of the Fermi-Dirac distribution function for the system. In the hot electron or quasi-equilibrium regime where the electronic temperature $T_e(r)$ is locally defined, the local effective electron temperature is given by

$$\Pi(r) = 2T_e(r) \tag{4.1.2}$$

and is proportional to the real local electron temperature. We define the local weighting function $g_{nm}(\mathrm{r}) = \nabla\phi_n \cdot \hat{\sigma}\nabla\phi_m$ so that we can rewrite the noise as

$$S_{nm} = 2\int d\boldsymbol{r} g_{nm}(\boldsymbol{r}) T_e(\boldsymbol{r}) \tag{4.1.3}$$

To obtain a measure of the local temperature, we must simplify the form of the local weighting function.

We first consider the simpler case of a rectangular geometry of length $L$ with two contacts spanning the width of the channel. The characteristic potentials are found by setting one terminal to $V_1 = 1$ and the other to $V_2 = 0$ and vice-versa, so that the two characteristic potentials are $\phi_1 = y/L$ and $\phi_2 = 1 - y/L$. Assuming that the conductivity is uniform and constant, the weighting function is then $g_{12}(\mathrm{r}) = -\sigma/L^2$, also a uniform constant, which is negative because the noise obeys the properties $\sum_n S_{nm} = 0$ and $S_{nn} > 0$ due to charge conservation. Taking the rectangle to have width $W$, we can express the noise in terms of the spatially-averaged temperature $T_e^{avg} = \int d\boldsymbol{r} T_e(\boldsymbol{r})/(L \times W)$, which results in $S_{12} = -2\sigma \frac{W}{L} T_e^{avg}$. The electrical conductance of the rectangle is $G = \sigma W/L$, so

$$S_{12} = -2\, G\, T_e^{avg}. \tag{4.1.4}$$

Here, $T_e^{avg} = T_{JN}$, the measured Johnson noise temperature[3,4]. This expression resembles the equilibrium multiterminal Johnson noise found by Sukhorukov and Loss for arbitrary geometry, $S_{nm} = -2\, G_{nm}\, T$, with the difference that Eq.4.1.4 holds for non-equilibrium temperature distributions.

To show how Eq.4.1.4 relates to the measurement, we observe that differential noise measures the difference in fluctuations between two terminals. This may be expressed as $\langle(\delta I_1 - \delta I_2)^2\rangle$, where $\delta I_{1,2}$ are the fluctuation currents of terminals 1 and 2. The corresponding noise correlator can be expressed by the difference of characteristic potentials, which we define as $S_{1-2,1-2} = 2\int d\boldsymbol{r} g_{1-2,1-2}(\boldsymbol{r}) T_e(\boldsymbol{r})$, where the differential weighting function is defined as $g_{1-2,1-2}(\mathrm{r}) = \nabla\frac{1}{2}(\phi_1 - \phi_2) \cdot \hat{\sigma}\nabla\frac{1}{2}(\phi_1 - \phi_2)$. Expanding, we find that $4S_{1-2,1-2} = S_{11} + S_{22} - S_{12} - S_{21}$. Invoking $\sum_n S_{nm} = 0$ and $S_{nm} = S_{mn}$, we find that $S_{1-2,1-2} = -S_{12}$. Thus, the two terminal differential noise is determined by Eq.4.1.4 and is proportional to $T_e^{avg}$.

Next, we examine how a multiterminal device modifies Eq.4.1.4. For simplicity, we first consider a 1D bridge located at the center of the hot and cold sides. From above, the differential correlator $S_{1-2,1-2} = 2\int d\boldsymbol{r} g_{1-2,1-2}(\boldsymbol{r}) T_e(\boldsymbol{r}) = 2\int d\boldsymbol{r} \nabla\frac{1}{2}(\phi_1 - \phi_2) \cdot \hat{\sigma}\nabla\frac{1}{2}(\phi_1 - \phi_2) T_e(\boldsymbol{r})$. To understand how the multiterminal geometry affects $S_{1-2,1-2}$, we thus must understand the difference of characteristic potentials of the measured terminals in different regions of the device. Due to symmetry of the device about the bridge axis, the characteristic potentials $\phi_1$ and $\phi_2$ are identical up to a reflection about the bridge axis. Therefore, their values in the bridge and hot regions are identical, and the integrand above vanishes everywhere except the cold side. The resulting $S_{1-2,1-2}$ is thus sensitive only to fluctuations on the cold side. Importantly, this conclusion holds regardless of whether the bridge is diffusive, since the characteristic potential gradient will vanish in a non-diffusive bridge.

In real devices, some corrections to the ideal case must be considered. In general, the device will not be perfectly symmetric about the bridge axis due to inevitable shifts during fabrication. These shifts arise due to imperfections in electron beam lithography, with errors of order 10-100nm, compared to the typical device dimensions of several microns. Although these are small, we proceed to consider their effect by defining terminals 1 and 2 as cold side top and bottom terminals, and 3 and 4 as hot side top and bottom terminals. Due to shifts, the bridge may be offset upward, toward terminals 1 and 3, or downwards, towards terminal 2 and 4. We rewrite the differential correlator in terms of the individual weighting functions, $S_{1-2,1-2} = 2\int dr g_{1-2,1-2}(r) T_e(r) = 2\int dr(g_{11} + g_{22} - g_{12} - g_{21}) T_e(r)$. If the bridge shifts upward, the bridge value of $g_{11}$ will grow, while that of $g_{22}$ will correspondingly shrink. On the other hand, $g_{12}$ will remain the same to first order, since the value of the weighting function at the bridge is determined by a product of two near-linear functions in the cold side, and so deviations are approximately quadratic about the center of the cold side. These deviations will thus tend to maintain the cancellation, yielding a negligible contribution outside of the cold region. In addition, real bridges may have a finite width, for which the corrections may be numerically shown to be vanishingly small, with hot and bridge contributions orders of magnitude smaller than the cold side contribution.

In sum, the combination of differential measurement and device geometry and symmetry ensures that the noise measurement of hot and cold sides corresponds to the local electronic temperature, as in Eq.4.1.4, which is borne out by the results shown in the main text.

## 4.2 The Cold Side as Power Meter

We now show how the measured temperature rise of the cold side, $\Delta T_C$, is related to the total heat flux across the bridge, $Q$. We first present the general result for any geometry, and then present a detailed calculation for the specific case of a rectangular noise thermometer.

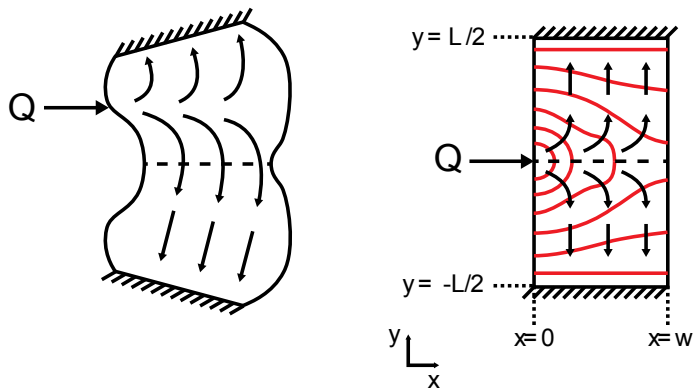

**Figure S4 Heating of the cold side by the bridge.** Left panel: energy current Q is injected at an arbitrary point along the boundary of a two-terminal geometry representing a conducting system with contacts at the bath temperature. The axis of symmetry is shown by the dashed line. Right panel: Rectangular case and geometric definitions used in section 4.2. Different conventions are used in section 7.

We consider a two-terminal cold side thermometer with contacts at the bath temperature and energy current $Q$ injected at some arbitrary point along its boundary (see Fig.S4). Here, only the energy current transmitted by the bridge is important, regardless of whether the bridge itself is diffusive or in a different state. As discussed in Sec.4.1, the Johnson noise temperature on the cold side $\Delta T_C$ is related to the current-current correlator $S$, which is determined by the conductivity, characteristic potentials and temperature distribution of the cold side. This relationship can be generalized, resulting in

$$\Delta T_C = R \int d^2r \, T(\vec{r})\vec{\nabla}\phi(\vec{r}) \cdot \left[\sigma\vec{\nabla}\phi(\vec{r})\right], \tag{4.2.1}$$

where $\phi(\mathbf{r})$ is the (dimensionless) characteristic potential discussed in the previous supplemental section for one of the two terminals (with subscripts dropped in the two-terminal setup), $R$ is the two-terminal electrical resistance between the contacts on the cold side, and the integral is taken over the entire area of the cold side. Manipulating this expression, and using the Wiedemann-Franz relation $\sigma = \kappa/(L_C T_0)$, where $L_C$ is the cold side Lorenz ratio and $T_0$ the bath temperature, as well as the continuity of the heat current $\vec{\nabla} \cdot (\kappa \vec{\nabla} T) = 0$, one can derive a general relation

$$\Delta T_C = \frac{6}{G_C^{th}} \int_C ds \, \phi(\vec{r})[1 - \phi(\vec{r})]\vec{q}(\vec{r}) \cdot (-\hat{n}). \tag{4.2.2}$$

In this expression, the integral is taken over the bounding contour $C$ of the cold side, and the quantity $\vec{q}(\vec{r}) \cdot (-\hat{n})$ represents the heat flux density entering the sample at a point $\vec{r}$ along the boundary. For the special case where the heat enters the sample at a point along an axis of bilateral symmetry halfway between the two contacts, $\phi = 1/2$, and one has $\Delta T_C = \frac{6}{G_C^{th}} \times \frac{1}{4} Q$, or

$$Q = \frac{2}{3} G_C^{th} \Delta T_C \tag{4.2.3}$$

A full derivation of Eq.4.2.2 will be presented in a forthcoming publication[4].

As an explicit example, we now derive the general result shown in Eq.4.2.3 for a rectangular geometry. We consider the heat flux as being injected at a point midway along the rectangular cold side, with two thermalizing contacts on top and bottom (see Fig.S4). Fourier's law holds that the heat flux density, $\vec{q}$, is related to the thermal conductivity $\kappa$, and the temperature distribution, $T$, by

$$\vec{q} = -\kappa \vec{\nabla} T \tag{4.2.4}$$

The total heat flux is related to the heat flux density by an integral across the cold side width, which by symmetry is

$$\frac{Q}{2} = \int_{x=0}^{x=w} dx \, \vec{q}(x,y) \cdot \hat{y} \tag{4.2.5}$$

Since we assume that the heat flux is conserved everywhere within the area of the sample (due to negligible energy loss to phonons), Eq.4.2.5 applies at every value of $y$. Inserting Fourier's law,

$$\frac{Q}{2} = -\kappa \int_{x=0}^{x=w} dx \, \frac{dT}{dy} \tag{4.2.6}$$

For simplicity, we can define $T$ relative to the bath temperature, so that

$$T\left(x, y = \pm \frac{L}{2}\right) = 0. \tag{4.2.7}$$

These boundary conditions allow us to relate the temperature distribution to its derivative via the integral $T(x,y) = -\int_{y'=0}^{y'=y} dy' \, \frac{dT}{dy'}$. Now, integrating the left and right sides of Eq.4.2.6,

$$\int_{y'=0}^{y'=y} dy' \frac{Q}{2} = -\kappa \int_{y'=0}^{y'=y} dy' \int_{x=0}^{x=w} dx \frac{dT}{dy'}, \qquad (4.2.8)$$

so that

$$\frac{Q}{2}y = \kappa \int_{x=0}^{x=w} dx\, T(x,y). \qquad (4.2.9)$$

Performing a second integration

$$\int_{y=0}^{y=\frac{L}{2}} dy\, \frac{Q}{2}y = \kappa \int_{y=0}^{y=\frac{L}{2}} dy \int_{x=0}^{x=w} dx\, T(x,y) \qquad (4.2.10)$$

The right side is the average of the temperature distribution, $\Delta T_C$, which is what is measured by the Johnson noise measurement in this geometry. Performing the integral on the left and rearranging therefore gives

$$Q = \frac{8W}{L}\kappa \Delta T_C. \qquad (4.2.11)$$

We can express $Q$ in terms of the thermal conductance $G_C^{th} = P_J^C/\Delta T_C^{s.h.}$ measured through self-heating, where $P_J^C$ is the Joule power applied to the cold side in self-heating and $\Delta T_C^{s.h.}$ is the cold side self-heating temperature rise. Using the self-heating result $\kappa = \frac{L}{12\,W} G_C^{th}$,[3,5] we obtain

$$Q = \frac{2}{3} G_C^{th} \Delta T_C. \qquad (4.2.12)$$

This coincides with the general result given in Eq.4.2.3.

# 5. Effective Thermal Circuit Connection

In this section, we will connect the real-space model of Fig.1c of the main text to the effective thermal circuit shown in Fig.1b of the main text. We will work in a simplified regime that will also demonstrate consistency with the general nonlocal power-temperature relationship derived in the previous section.

We begin with the geometry shown in Fig.S5. In the figure, all regions are assumed to be diffusive, whereas for the analysis below, only the hot and cold sides are required to be diffusive, while the bridge need not be. For a sufficiently narrow bridge and cold side, we can use the simplification that all injected Joule power is dissipated in the hot side. Under these conditions, the hot side temperature distribution is parabolic, the bridge possesses a linear temperature profile, and the cold side a Λ-shaped temperature profile peaked at the cold end of the bridge and terminating at the bath temperature of the contacts.

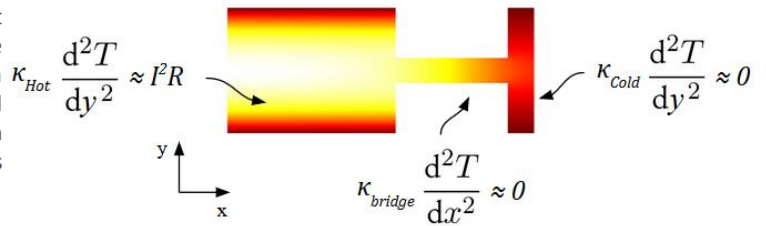

**Figure S5 Device temperature distribution in the diffusive regime.** Central image: finite element simulation of the temperature distribution (see Fig1c of the main text for details). Heat equation approximations for the hot, bridge, and cold regions are shown, assuming a diffusive regime in all regions, where the bridge diffusive regime is not required for the calculation.

We now seek to relate this real-space model to the effective thermal circuit shown in Fig.1b of the main text. From that model, we found $G_{bridge}^{th} = G_{C\prime}^{th} \Delta T_{C\prime} / (\Delta T_{H\prime} - \Delta T_{C\prime})$ and $Q_{bridge} = G_{C\prime}^{th} \Delta T_{C\prime}$, with the primed quantities corresponding to those of the effective thermal circuit. In the effective thermal circuit, $\Delta T_{C\prime}$ is related to the temperature at the cold end of the bridge. In the real space model, this is $\Delta T_C^{peak}$, the difference between the peak temperature of the cold side at the cold end of the bridge and the bath. We thus have that

$$Q_{bridge} = \Delta T_C^{peak} G_{C\prime}^{th}. \tag{5.1}$$

Now assume a rectangular cold side shape of length $L$ and width $W$, with a linearly-varying temperature distribution. We can then compute $G_{C\prime}^{th}$ in terms of $\kappa_{cold}$, the cold side thermal conductivity,

$$G_{C\prime}^{th} = \kappa_{cold} \frac{W}{L/2} \times 2. \tag{5.2}$$

Here, we treat the cold side as two parallel channels of length $L/2$ conducting heat to the thermal bath at the electrical contacts. We now use the self-heating result[3]

$$\kappa_{cold} = \frac{P_J^C}{T_{JN}^{CC}} \frac{L}{W} \frac{1}{12}, \tag{5.3}$$

where $P_J^C$ is the Joule power applied to the cold side, and $T_{JN}^{CC}$ is the self-heating temperature rise measured on the cold side, which results in

$$Q_{bridge} = \Delta T_C^{peak} \frac{P_J^C}{T_{JN}^{CC}} \frac{1}{3}, \tag{5.4}$$

from which we can make the identification

$$G_{C\prime}^{th} = G_C^{th,s.h.} \frac{1}{3}. \tag{5.5}$$

Equation 5.6 shows that the effective thermal circuit quantity $G_{C\prime}^{th}$ can be directly related to the measured self-heating value $G_C^{th,s.h.}$ through a simple numerical factor. This is the main result of this section.

We now demonstrate consistency with the energy current relation 4.2.3. First, we express $\Delta T_C^{peak}$ in terms of the measured $\Delta T_C = \int_C d\mathbf{r}\, T_e(\mathbf{r})/(L \times W)$, the Johnson noise average over the cold side. The path $C$ goes between the two opposite electrodes at the bath temperature, crossing at its midpoint the bridge location where the temperature is maximal. Since the temperature distribution is linear in the absence of energy loss to phonons, the integration is simple and we have $\Delta T_C^{peak} = 2\Delta T_C$. Putting these two results together we obtain $Q_{bridge} = \Delta T_C^{peak} \, G_{Cl}^{th} = \frac{2}{3} \Delta T_C \, G_C^{th,s.h.}$, which coincides with the general result Eq.4.2.3.

# 6. Nonlocal Thermometry with Electron-Phonon Coupling

Here, we consider how electron-phonon coupling in the nonlocal noise thermometers modifies the thermal transport analysis (electron-phonon coupling in the bridge is considered in Supplementary Sections 13 and 14). In graphene, the electronic diffusion cooling and electron-phonon energy loss terms can be individually, quantitatively determined[5–7], enabling the nonlocal measurement even in the presence of electron-phonon coupling, as we show below.

Extracting the thermal conductance $G_{bridge}^{th}$ of the bridge requires two separate measurements:

1. Running a current between the contacts of the cold side and measuring the increase of Johnson noise temperature $\Delta T_{JN}^{CC}$ on the cold side
2. Running a current between the contacts of the hot side and measuring the increase in Johnson noise temperature $\Delta T_{JN}^{HC}$ on the cold side

We consider these two situations one at a time, taking into account the effects of electron-phonon coupling. First, we outline the general structure of the heat equation that governs the spatial profile $T(r)$ of the electron temperature

If the thermal conductivity $\kappa$ is taken to be constant and independent of position $r$, then the heat equation is

$$-\kappa \nabla^2 T(r) + \Sigma_{eph}\left[T(r)^\delta - T_0^{\,\delta}\right] = p(r) \tag{6.1}$$

Where $\Sigma_{eph}$ is the electron-phonon coupling constant, $\delta$ is an exponent characteristic of the system dimensionality and energy loss mechanism, $T_0$ is the bath temperature of the phonons and electrons in the boundary reservoirs, and $p(r)$ is the dissipated power density added to the electron system at point $r$. Throughout this section we work in the limit where the heating is weak enough that $T - T_0 \ll T_0$, so that equation 6.1 can be linearized, $T(r)^\delta - T_0^{\,\delta} \approx \delta T_0^{\,\delta-1}(T - T_0)$. It is convenient to change notation so that $T$ denotes the temperature relative to the base temperature, $T - T_0 \rightarrow T$. With these simplifications the heat equation is

$$\nabla^2 T(r) - \frac{T(r)}{L_{eph}^{\,2}} = -\frac{p(r)}{\kappa} \tag{6.2}$$

Here we have introduced the electron-phonon coupling length,

$$L_{eph} = \sqrt{\frac{\kappa}{\delta \Sigma_{eph} T_0^{\,\delta-1}}} \tag{6.3}$$

whose parameters may be extracted from experiments[5–7].

## Temperature profile from self-heating

We first consider measurement 1, where the cold side experiences Joule heating due to an electrical current between the two cold side contacts. Because of the balanced circuit shown in Fig.1 of the main text, the current flows essentially uniformly across the cold side. This would also be the case if the hot side terminals were both electrically floating, or if the bridge resistance was much greater than the cold side resistance. Under at least one of these conditions, the Joule power is uniform in space and

$$p(r) = \frac{P_J^C}{W\,L} \qquad (6.4)$$

Where $L$ is the length of the cold side (the distance between the two contacts), $W$ is the width, and $P_J^C = I_{cold}{}^2 R_{cold}$ is the Joule power dissipated when there is a current $I_{cold}$ and resistance $R_{cold}$.

We define the two contacts to be at $y = 0$ and $y = L$. The contacts are assumed to be perfect heat sinks so that $T(0) = T(L) = 0$. One can then solve the equation 6.2 to get

$$T(y) = \frac{P_J^C}{\kappa_C}\frac{L_{eph}{}^2}{L\,W}\left[1 - \cosh\left(\frac{L-2y}{2L_{eph}}\right)\text{sech}\left(\frac{L}{2L_{eph}}\right)\right] \qquad (6.5)$$

In the limit where there is no electron-phonon coupling, $L_{eph} \to \infty$, this equation becomes the familiar parabolic temperature profile:

$$T(y) \to \frac{1}{2}\frac{P_J^C}{\kappa_C}\frac{y(L-y)}{L\,W}. \qquad (6.6)$$

From equation 6.5 one can read off the temperature at any given point, for example the maximum at $y = L/2$

$$T_{max}(y) = T\left(y = \frac{L}{2}\right) = \frac{P_J^C}{\kappa_C}\frac{L_{eph}{}^2}{L\,W}\left[1 - \text{sech}\left(\frac{L}{2L_{eph}}\right)\right]. \qquad (6.7)$$

One can also calculate the Johnson noise temperature, which in the rectangular geometry is a simple average of the temperature profile, $T_{JN} = (1/L)\int_0^L T(y)dy$. This gives

$$T_{JN} = \frac{P_J^C}{\kappa_C}\frac{L_{eph}{}^2}{L\,W}\left[1 - 2\frac{L_{eph}}{L}\tanh\left(\frac{L}{2L_{eph}}\right)\right]. \qquad (6.8)$$

Now we can determine the thermal conductivity $\kappa_C$ by measuring the Johnson noise temperature

$$\kappa_C = \frac{P_J^C}{\Delta T_{JN}^{CC}}\frac{L_{eph}{}^2}{L\,W}\left[1 - 2\frac{L_{eph}}{L}\tanh\left(\frac{L}{2L_{eph}}\right)\right] \qquad (6.9)$$

## The temperature profile from heat coming across the bridge

Now we consider the second measurement, where current is applied between the two contacts of the hot side, and all the heating of the cold side is provided by thermal energy current conducted across the bridge. This means that the dissipating Joule power $p(r)$ is equal to zero everywhere except at the hot side.

One can again solve equation 6.2 for the temperature profile $T(r)$. This solution becomes particularly simple when $L_{eph} \ll W$, i.e. when the cold side is much narrower than the electron-phonon coupling length. In this case $T(r)$ effectively depends only on the vertical position $y$, and

$$T(y) = T_C \sinh\left(\frac{y}{L_{eph}}\right) \text{csch}\left(\frac{L}{2L_{eph}}\right) \tag{6.10}$$

Where $T_C = T(y = L/2)$ denotes the maximum temperature on the cold side, right at the end of the bridge. The corresponding Johnson noise temperature

$$T_{JN} = \frac{1}{2} T_C \frac{4L_{eph}}{L} \tanh\left(\frac{L}{4L_{eph}}\right) \tag{6.11}$$

In the limit where the electron-phonon coupling is turned off, $L_{eph}/L \to \infty$, these two equations give the familiar linear temperature profile $T(y) = T_C \, y/(L/2)$ and $T_{JN} = T_C/2$.

The value of the temperature $T_C$ is determined by the total heat $Q$ flowing across the bridge. In particular,

$$Q = 2\kappa W \left|\frac{dT}{dy}\right|_{y=\frac{L}{2}} \tag{6.12}$$

That is, the slope of the temperature profile gives the magnitude of the heat current. Evaluating this expression, and plugging in the expression for $T_{JN}$, gives

$$Q = \Delta T_{JN}^{HC} \kappa \frac{L\,W}{L_{eph}^2} \coth\left(\frac{L}{4L_{eph}}\right) \coth\left(\frac{L}{2L_{eph}}\right) \tag{6.13}$$

In the limit of $L_{eph}/L \to \infty$, this becomes the familiar

$$Q = \Delta T_{JN}^{HC} \kappa \frac{8W}{L} \tag{6.14}$$

**Putting it together to get $G_{bridge}^{th}$**

We want to estimate the thermal conductance of the bridge $G_{bridge}^{th}$ defined by

$$G_{bridge}^{th} = \frac{Q}{T_H - T_C} \tag{6.15}$$

All of the necessary ingredients for estimating $T_H$, $T_C$, and $Q$ have been outlined in the above [Eq.6.7 and 6.14], provided that we know the electron-phonon coupling strength, which is possible to independently obtain *in situ* in graphene devices with thermal noise measurements[6–8].

We now seek to understand what effect $L_{eph}$ has on the resulting value of $G_{bridge}^{th}$. To do so, we expand the expression in Eq.6.15 in orders of $L_{eph}$. Up to the first correction, we have

$$G_{bridge}^{th} = \frac{\frac{2}{3}\Delta T_C\, G_C^{th}}{\frac{3}{2}\Delta T_H - 2\,\Delta T_C} + \frac{L^2}{L_{eph}^2}\frac{\Delta T_C\, G_C^{th}(6\Delta T_H + 16\Delta T_C)}{180(3\Delta T_H + 4\Delta T_C)^2} + O\left(\frac{L^3}{L_{eph}^3}\right), \qquad (6.16)$$

where we have used $L = L_H = L_C$ as the length of the hot and cold sides, $\Delta T_H$ is the measured Johnson noise temperature of the hot side upon Joule heating ($\Delta T_{JN}^{HH}$ in the notation above), and $\Delta T_C$ is the measured Johnson noise temperature of the cold side when the hot side is heated (equivalent to $\Delta T_{JN}^{HC}$ above). The first term on the right-hand side of Eq.6.16 is the result obtained when $L_{eph} \to \infty$, described in previous sections. The first correction is positive, so that neglecting electron-phonon coupling when it is present in the noise thermometers leads to an underestimate of $G_{bridge}^{th}$. The scaling with $(L/L_{eph})^2$ and the large numerical coefficient of 180 in the denominator of Eq.6.16 suggests that this correction may be small.

# 7. Excess Energy Loss in the Nanotube-Graphene Devices

In this section, we discuss the presence of excess energy loss in the nanotube-graphene devices. These devices are made without a top hBN layer in order to enable electrical contact between the graphene and nanotube. Due to the lack of top hBN, additional disorder is present in the graphene thermometers,

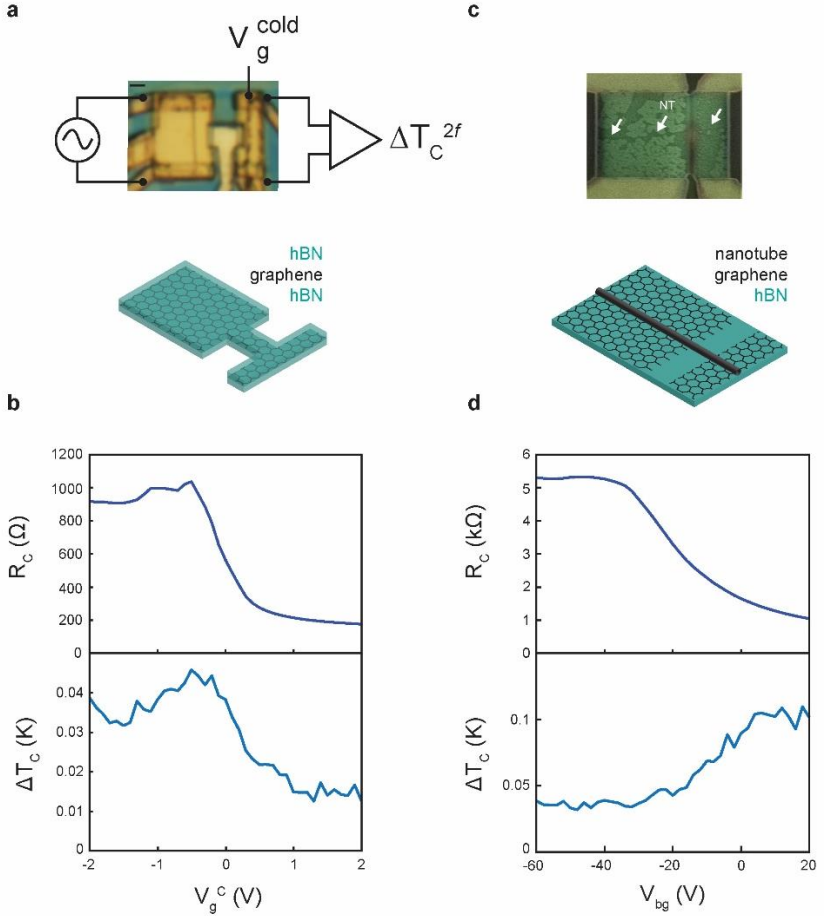

**Figure S6 Cold side resistance and temperature change versus gate voltage. a**) Top panel: circuit schematic for the displayed measurements. Joule power is applied to the hot side and temperature change is measured on the cold side as a function of the local cold side top gate voltage $V_g^{cold}$. Bottom panel: device stack (see Fig.1d and captions for details). **b**) Cold side graphene resistance $R_C$ and temperature change upon hot side heating $\Delta T_C$ versus local cold side topgate voltage $V_g^C$. **c**) Top panel: composite optical and scanning electron micrograph of the CNT bridge device (see Fig.3a and caption for details). Bottom panel: device stack. **d**) Cold side graphene resistance $R_C$ and temperature change upon hot side heating $\Delta T_C$ versus global backgate voltage $V_{bg}$.

due to fabrication residue and topographic corrugation.

We demonstrate this energy loss by sweeping the cold side gate, as shown in Fig.S6. Power is applied to the hot side and $\Delta T_C$ is measured as the cold side gate is swept, changing the cold side resistance. In the fully hBN-encapsulated graphene device, the cold side Dirac peak is visible in Fig.S6b, top panel. The corresponding $\Delta T_C$, Fig.S6b, bottom panel, evolves in correlation with the cold side resistance. This is the behavior expected from the thermal circuit shown in Fig.1b of the main text. For this thermal circuit, one can show that $\Delta T_C = Q_{in}\, G^{th}_{bridge}/\sum_{ij} G^{th}_i G^{th}_j$. As $G^{th}_C$ increases, $\Delta T_C$ decreases, as expected.

In the nanotube-graphene device, however, the behavior is markedly different. At the height of the Dirac peak (Fig.S6d, top panel), $\Delta T_C$ is at a minimum and increases as the resistance decreases. This suggests that an additional energy loss mechanism is present, which may be maximal near charge neutrality where it acts to suppress $\Delta T_C$ despite the decreased cold side thermal conductance. Measurements shown in the main text are taken near charge neutrality of the cold side in order to minimize the cold side thermal conductance and thus maximize its sensitivity to the small thermal conductance of the nanotubes. Therefore, we interpret the measured electronic thermal conductance of the nanotubes shown in the main text as lower bounds, due to the excess heat loss present at charge neutrality of the cold side graphene.

Past work has demonstrated that disordered graphene possesses larger energy loss compared to fully-encapsulated devices[5,7,9,10]. This is believed to be due to enhanced electron-phonon interaction in these devices (although device size, which also determines electron-phonon coupling, complicates this understanding). Further effects may arise from the corrugations of partially- or non-encapsulated graphene, which are predicted to lead to an energy loss mechanism due to scattering off of pinned flexural phonons[11–15]. Future development of nanotube-graphene devices with fully-encapsulated graphene may mitigate these issues.

## 8. Theory of Plasmon Hopping Energy Transport

Here we obtain an expression for the energy current in a one-dimensional conductor with long-range interactions between electrons in the presence of a single impurity that blocks transmission of particles. For simplicity, we start with the model of spinless electrons; the effect of the spin and valley degrees of freedom will be discussed later. We describe the electrons in the conductor as a Luttinger liquid[16] with the Hamiltonian $\widehat{H}_0 + \widehat{V}$, where

$$\widehat{H}_0 = \frac{\hbar v}{2\pi} \int_{-\infty}^{\infty} \left[ K(\partial_x \theta)^2 + \frac{1}{K}(\partial_x \phi)^2 \right] dx, \quad \phi(0) = 0, \quad \partial_x \theta(0) = 0 \tag{8.1}$$

$$\widehat{V} = \frac{1}{\pi^2} \int_0^{\infty} dx \int_{-\infty}^{0} dy\, V(x-y) \partial_x \phi(x) \partial_y \phi(y) \tag{8.2}$$

Here $v$ is the velocity of the bosonic excitations of the liquid (plasmons), $K$ is the Luttinger liquid parameter, while $\phi(x)$ and $\theta(x)$ are the standard bosonic fields with the commutation relation

$$[\phi(x), \partial_y \theta(y)] = i\pi \delta(x-y) \tag{8.3}$$

Their physical meaning is that the electron density is given by $n(x) = n_0 + \partial_x \phi(x)/\pi$, while the momentum per particle of the Luttinger liquid is $\kappa(x) = -\hbar \partial_x \theta(x)$, where $n_0$ is the mean electron density. The boundary conditions in Eq.8.1 account for a strong point-like impurity at $x = 0$ that completely pins the Luttinger liquid and does not allow transport of electrons through it. The two parts of the system described by the Hamiltonian in Eq.8.1 corresponding to positive and negative $x$ are completely decoupled from each other. In the full Hamiltonian the coupling term $\hat{V}$ is due to the long-range tail of the interaction potential $V(x)$ describing the interaction between electrons.

For the subsequent discussion, it is convenient to express the bosonic fields in terms of the creation and annihilation operators $b_k$ and $b_k^\dagger$. Unlike the standard Luttinger liquid theory, this relation should account for the barrier at $x = 0$. It is convenient to denote the wavevector of the boson by $k$ for the liquid to the left of the barrier and by $q$ for that to the right of the barrier. Then the bosonic fields take the form

$$\phi(x) = \vartheta(-x)\sqrt{K}\int_0^\infty \frac{dk}{\sqrt{k}}(b_k + b_k^\dagger)\sin(kx) + \vartheta(x)\sqrt{K}\int_0^\infty \frac{dq}{\sqrt{q}}(b_q + b_q^\dagger)\sin(qx),$$

$$\partial_x \theta(x) = -\vartheta(-x)\frac{i}{\sqrt{K}}\int_0^\infty dk\,\sqrt{k}(b_k - b_k^\dagger)\sin(kx) - \vartheta(x)\frac{i}{\sqrt{K}}\int_0^\infty dq\,\sqrt{q}(b_q - b_q^\dagger)\sin(qx),$$
(8.4)

where $\vartheta(x)$ is the unit step function. Taking into account the commutation relations $[b_k, b_{k'}^\dagger] = \delta(k-k')$, $[b_q, b_{q'}^\dagger] = \delta(q-q')$, and $[b_k, b_q^\dagger] = 0$ for the bosonic operators, one can verify that the commutation relation 8.3 for $\phi$ and $\theta$ is satisfied. Substitution of Eqs.8.4 into Eq.8.1 yields

$$H_0 = \hbar v \int_0^\infty dk\, k b_k^\dagger b_k + \hbar v \int_0^\infty dq\, q b_q^\dagger b_q, \tag{8.5}$$

where we omitted the constant corresponding to the zero-point energy. Predictably, the excitations of the Luttinger liquid in the presence of the barrier at $x = 0$ are two sets of plasmons with energies $\epsilon_k = \hbar v k$ and $\epsilon_q = \hbar v q$.

Substitution of Eqs.8.4 into Eq.8.2 expresses the operator $\hat{V}$ in terms of the bosonic operators,

$$\hat{V} = \int_0^\infty dq \int_0^\infty dk\, t_{qk}(b_q + b_q^\dagger)(b_k + b_k^\dagger), \tag{8.6}$$

where

$$t_{qk} = \frac{K}{\pi^2}\frac{\sqrt{qk}}{q^2 - k^2}\int_0^\infty V(x)[q\sin(qx) - k\sin(kx)]dx. \tag{8.7}$$

To make further progress, one should specify the form of interaction $V(x)$ between the electrons. In the case of a nanotube in the presence of a gate at a distance $d$ from it, due to the resulting image charge $V(x)$ has the form

$$V(x) = \frac{e^2}{|x|} - \frac{e^2}{\sqrt{x^2 + 4d^2}}. \tag{8.8}$$

For small values of the wavevectors, Eq.8.7 yields

$$t_{qk} = \frac{2K}{\pi^2} e^2 d \sqrt{qk}, \qquad q, k \ll \frac{1}{d}. \tag{8.9}$$

In the opposite limit, one can neglect the second term in Eq.8.8, which reduces it to pure Coulomb repulsion. In this case,

$$t_{qk} = \frac{K}{2\pi} e^2 \frac{\sqrt{qk}}{q+k}, \qquad q, k \gg \frac{1}{d}. \tag{8.10}$$

We are now in a position to study energy transport through the impenetrable barrier. We assume that the left and right subsystems are in thermal equilibrium states with different temperatures $T_L$ and $T_R$. This implies that the occupation numbers of the plasmons in both subsystems take the form of Bose distributions,

$$N_k^L = \frac{1}{e^{\hbar v k/T_L} - 1}, \quad N_q^R = \frac{1}{e^{\hbar v q/T_R} - 1}. \tag{8.11}$$

The coupling of the left and right subsystems described by Eq.8.6 contains terms proportional to $b_q^\dagger b_k$ and $b_k^\dagger b_q$, which describe hopping of plasmons through the barrier. The rate of energy transport from left to right $J_E = -dE_L/dt$ is easily obtained from Fermi's golden rule,

$$J_E = \frac{2\pi}{\hbar} \int_0^\infty dk \int_0^\infty dq \, |t_{qk}|^2 \delta(\epsilon_k - \epsilon_q) \epsilon_k [N_k^L(N_q^R + 1) - N_q^R(N_k^L + 1)]. \tag{8.12}$$

After straightforward manipulations this expression takes the form

$$J_E = j_E(T_L) - j_E(T_R), \tag{8.13}$$

where

$$j_E(T) = \frac{2\pi}{\hbar} \int_0^\infty |t_{qq}|^2 \frac{q \, dq}{e^{\hbar v q/T} - 1}. \tag{8.14}$$

At low temperatures $T \ll \hbar v/d$ the integral in Eq.8.14 is dominated by small values of $q$, and one can use the expression in Eq.8.9 for the matrix element $t_{qk}$. This results in

$$j_E(T) = \frac{8\pi}{15} \frac{K^2 e^4 d^2}{\hbar^5 v^4} T^4. \tag{8.15}$$

At higher temperatures, $T \gg \hbar v/d$, one should use the expression 8.10 for $t_{qk}$, which yields

$$j_E(T) = \frac{\pi}{48} \frac{K^2 e^4}{\hbar^3 v^2} T^2. \tag{8.16}$$

We therefore conclude that in the case of pure Coulomb repulsion the energy current through the barrier $J_E$ is proportional to $T_L^2 - T_R^2$, whereas in the presence of a screening gate $J_E \propto T_L^4 - T_R^4$.

In the case of pure Coulomb interactions, the above conclusion is not entirely accurate. Due to the long-range nature of the interaction potential, the Luttinger liquid parameter $K$ is a weak function of temperature,

$$K = \frac{1}{\sqrt{1 + \frac{2e^2}{\pi \hbar v_F} \ln \frac{\hbar v}{wT}}}. \tag{8.17}$$

Here $w$ is a short-distance cutoff of the order of the width of the one-dimensional channel. The additional logarithmic temperature dependence can be ignored when comparing the theoretical expression for the energy current given by Eqs.8.13 and 8.16 with experimental data.

In the case of a one-dimensional system with spin and/or valley degrees of freedom, the above calculation requires some modification. One should start by bosonizing each channel and introducing orthogonal linear combinations of bosonic fields, with one of them, $\phi(x) = \sum_i \phi_i(x)/\sqrt{N}$, describing excitations of the charge density. [Here $N$ is the total number of channels; $N = 4$ for the electron system in a nanotube due to the spin and valley degrees of freedom. The normalization factor is required in order to preserve the commutation relations Eq.8.3.] This is the plasmon mode, which is described by the same Hamiltonian $H_0$ given by Eq.8.1, but the expression 8.2 for the coupling term $\hat{V}$ requires modification. The charge density $n(x) = n_0 + \sum_i \partial_x \phi_i(x)/\pi = n_0 + \sqrt{N} \partial_x \phi(x)/\pi$. Thus, the coupling term 8.2 acquires an additional factor $N$, and our results 8.15 and 8.16 should be multiplied by $N^2$.

## 9. Multi-thermometer Device Design

The results presented in the main text are from two-thermometer H devices, which measure thermal conductance. To realize a measurement of thermal conductivity, more thermometers are required. In this section, we present a possible design of a multi-thermometer device which shows how multiple graphene thermometers may be incorporated for future measurements of thermal conductivity.

Our method is based on two-terminal graphene rectangles as electronic noise thermometers. As such, a multi-thermometer device requires the bridge to be connected to several graphene rectangles, each possessing a local differential noise measurement. An example of such a design is shown in Figure S7. Here, the hot side on the left serves as the source of Joule heat. On the far right, the cold end is measured as shown in the main text. On the top and bottom edges of the bridge, four legs connect the bridge to four rectangular regions, each possessing a contact pair with a differential noise amplification chain. A finite-element simulation of the temperature distribution given that the hot side is Joule heated and all other terminals are electrically floating, assuming a bath temperature at the contacts, no thermal contact resistance, uniform conductivity, and diffusive behavior throughout all regions. This design, with four thermometers flanking the bridge, enables a thermal Hall effect measurement. For accurate measurement, the bridge and thermometer connections should be designed such that the heat flow is mainly across the bridge. If there is non-negligible heat flow to the thermometers, analysis may be performed to extract the bridge contribution and contact contributions as demonstrated by past work on resistive thermometry using multiple thermometers[17].

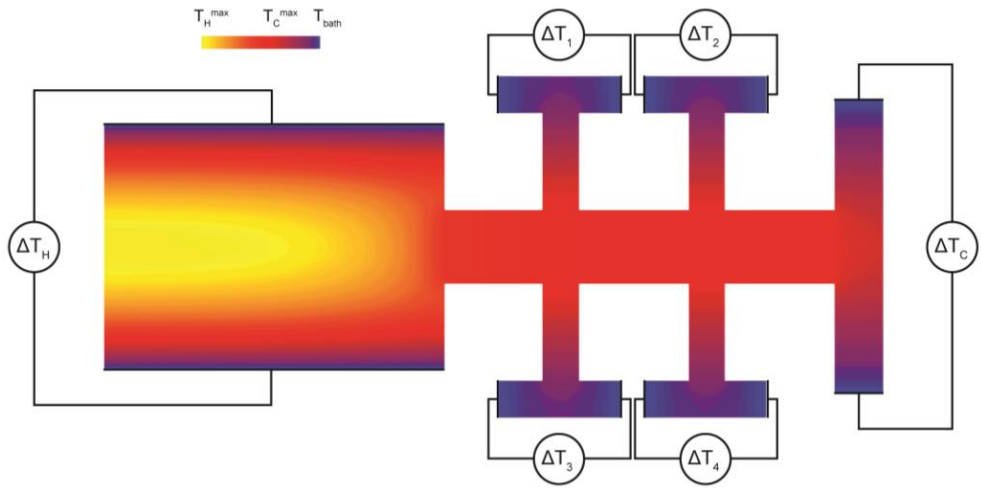

*Figure S7 Multi-thermometer nonlocal noise device for thermal conductivity and thermal Hall measurements. Shown is a finite element simulation of the temperature distribution assuming Joule heating of the hot side and all other terminals electrically floating. All terminals sit at a fixed bath temperature, and a uniform conductivity, diffusive behavior, and no thermal contact resistance is assumed.*

## 10. Nonlocal Thermometry Circuit

In the main text, we presented a simplified schematic of the measurement circuit (see Fig.1c). In this section, we present a detailed circuit diagram, shown in Fig.S8.

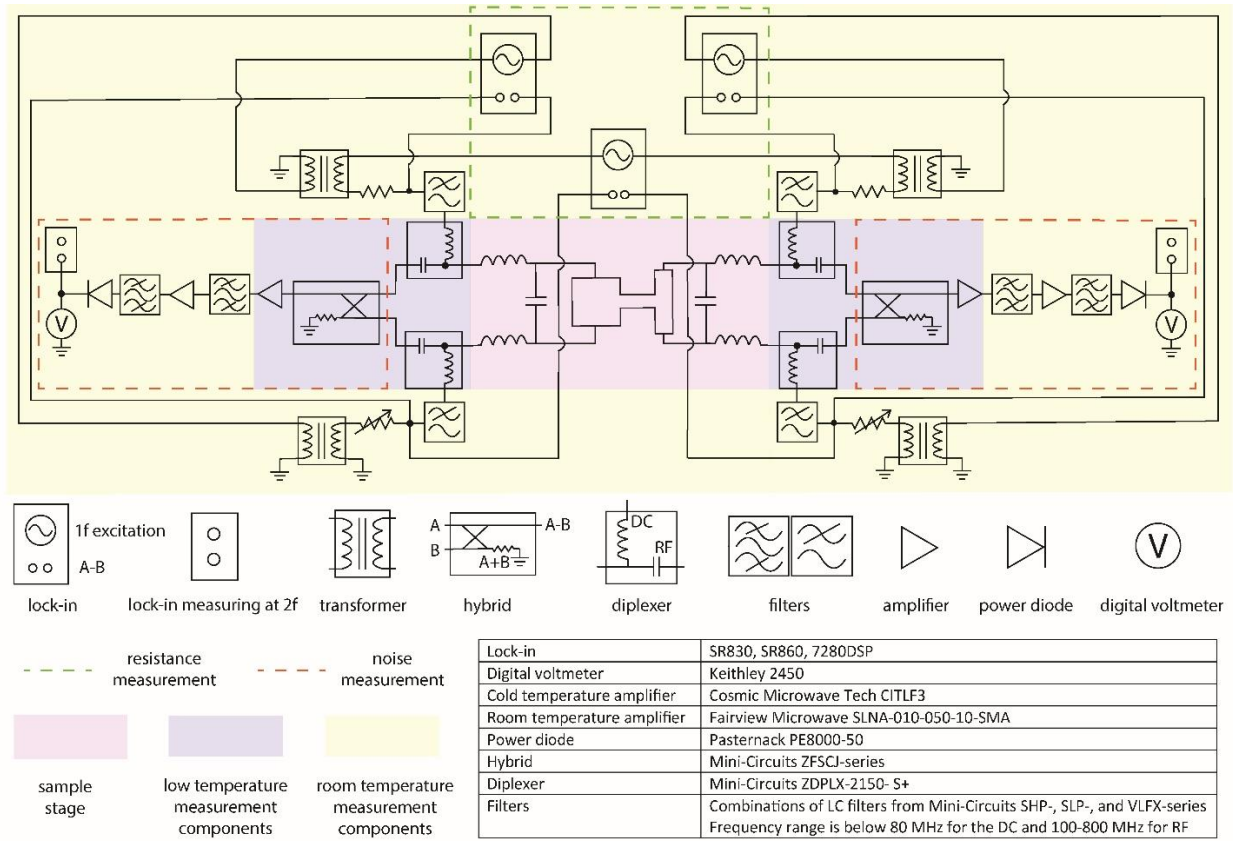

*Figure S8 Detailed circuit schematic for nonlocal noise thermometry with concurrent four-terminal bridge electrical measurement. A simultaneous hot, cold, and bridge electrical and thermal measurement circuit is shown. Hot, cold, and bridge lock-in amplifiers (green box) are operated at different frequencies. A four-terminal bridge electrical measurement is shown. The two differential matching circuits are chosen to have center frequencies spaced by at minimum the circuit bandwidths (see discussion).*

The sample, represented by the H device in the central pink region, is connected to the two differential matching circuits for hot and cold side noise thermometry. These are mounted on a sample stage with temperature sweeping capability, to define bath temperatures and perform calibration (see Supplementary Section 1). The two matching circuits are chosen to have non-overlapping bandwidths in frequency space, to ensure that each side measures its own local noise without crossover signals.

Flanking the matching circuits are two purple regions representing low-temperature measurement components. These components include diplexers to enable separate low-frequency biasing and measurement for electrical transport, and high-frequency noise measurement via a 180-degree hybrid and cryogenic low noise amplifier (LNA).

The two noise measurement amplification chains are highlighted by the dashed red boxes. After the cryogenic LNA, the noise signal is bandpass filtered, amplified by a high dynamic range amplifier at room temperature, bandpass filtered again, and fed into a power diode. This generates an output voltage proportional to the RF power integrated over a frequency bandwidth. The output voltage contains a DC component proportional to the total noise, which is measured by a digital voltmeter and used for calibration (as described in Supplementary Section 1). During the nonlocal noise measurement, the output voltage also contains a $2f$ component due to the low-frequency (10-100 Hz) $1f$ voltage excitation applied to the device, leading to $2f$ Joule power dissipation on the hot side. The diode output is thus fed into a lock-in amplifier set to the $2f$ heating frequency. This is repeated on the two sides, with different $1f$ signal frequencies used on hot and cold sides, and different bandpass filters chosen for the two different frequency bandwidths of the hot and cold matching circuits. Alternatively, the $2f$ heating experiment may be performed with the cold side floating, and its thermal conductance measured separately, as was done for the graphene bridge measurements in Fig.2 of the main text.

The green dashed line box (top of Fig.S8) shows the three low-frequency lock-ins for hot, cold, and bridge electrical resistance measurement. The hot and cold sides are measured in a two-terminal, current-biased configuration. The bridge measurement may be performed in several modes; Fig.S8 shows a four-terminal bridge measurement. As such, the lock-in excitation for the bridge is in a current biased mode and added to the hot and cold sides with a pair of transformers. The graphene bridge electrical measurements shown in Fig.2 of the main text are performed in four-terminal mode. Electrical measurements of the nanotube bridge shown in Fig.3 are performed in a two-terminal, voltage-biased mode.

## 11. Considering Contact Resistance

In this section, we consider how contact resistance, either at the metal-to-graphene interface or the graphene-to-bridge interface, affects our measurements.

For the hot and cold sides, the low contact resistance possible with 1-dimensional contacts of 2-dimensional graphene (see Methods reference 1 of the main text) allows for accurate self-heating measurements of the noise thermometers, as was previously demonstrated (see main text references 7, 23, 24, 34 and 37). Past work has considered the effect of contact resistance on self-heating measurements[18].

For the bridge, both electrical and thermal contact resistance, $R_C$ and $R_C^{th}$, respectively, can affect the measured Lorenz ratio, $L_{meas}/L_0 = (R_{bridge} + R_C)/L_0 T(R_{bridge}^{th} + R_C^{th})$, where $R_{bridge}$ and $R_{bridge}^{th}$ are the electrical and thermal resistances of the bridge. We do not have an exact estimate of $R_C$ and $R_C^{th}$ in our measurement (although extending our experimental technique will allow us to quantify contact contributions as discussed in Supplementary Section 9). However, for the graphene bridge (data shown in Fig.2 of the main text), there are several signs that contact contributions are negligible. First, since the device is monolithically fabricated out of a single, seamless sheet of graphene, we expect that $R_{bridge} \gg R_C$. Our measurement shows that $L_{meas}/L_0 \sim 1$, showing small variations with gate voltage and temperature. Since $R_{bridge}$ varies over a large range over the same gate voltage and temperature variation, we thus conclude that $R_{bridge}^{th} \gg R_C^{th}$ and $L_{meas}/L_0 \sim 1 \sim R_{bridge}/L_0 T R_{bridge}^{th}$. Second, our measured $L_{meas}/L_0$ shows a non-monotonic behavior with gate voltage at elevated temperatures, exhibiting a marked reduction from 1 in the low density regime. This observation is consistent with expectations from hydrodynamic theories of electronic energy transport in graphene channels (see main text, page 7). This observation also suggests that $R_{bridge}^{th} \gg R_C^{th}$, since such gate modulations would not be observed if contact resistance dominated.

For the nanotube devices shown in Figure 3, contact resistance may play a more important role. However, similar to the graphene case, the clear gate dependence of the thermal and electrical conductances and the Lorenz ratio suggests that other physics is at play, such as the plasmon hopping we propose. Nonetheless, we proceed to analyze in more detail the effect of contact resistance.

In 1D conductors, the known presence of quantum contact resistance can dominate, such as in the ballistic regime. Measurements of 1D channels have shown that ballistic quantum channels possess a quantum of thermal conductance with a Lorenz ratio of $1^{19-22}$. We can express this as $L_{contact}/L_0 = 1 = R_C / L_0 T R_C^{th}$. If the ballistic channel possesses an extrinsic (non-quantum) contact resistance, or a bulk channel resistance, we can define the Lorenz ratio of this contribution as $L_d/L_0 = R_d / L_0 T R_d^{th}$. The total measured Lorenz ratio will then be $L_{meas}/L_0 = (R_d + R_C)/L_0 T(R_d^{th} + R_C^{th}) = (1 + R_d/R_C)/(1 + L_0/L_d \times R_d/R_C)$. For strong Lorenz suppressions such that $L_d \ll L_0$, the device contribution bottlenecks the heat flow, and we have that $L_{meas}/L_0 \propto L_d/L_0$. However, for Lorenz enhancements $L_d \gg L_0$, we have that the contact contribution dominates, and the device Lorenz ratio becomes a small correction to the measured value, $L_{meas}/L_0 \approx (1 + R_d/R_C) \times (1 - L_0/L_d \times R_d/R_C)$. In this case, the leading contribution is the constant $1 + R_d/R_C$.

Using the formulas above, we can now estimate this number noting that carbon nanotubes possess a degeneracy of 4 for spin and valley isospin, giving a quantum contact resistance of $\sim 6.4 k\Omega$. Taking the more conductive Device 1, at its most conducting point at T=70K, the channel resistance is $\sim 20 k\Omega$, from which we can estimate $R_d/R_C \sim 2.1$, so that the maximum measured Lorenz value would be $L_{meas}/L_0 \approx 3$. This value is exceeded in Device 1, particularly at low electron density close to the gap, and at lower temperature throughout the density range (see Supplementary Section 12). We can therefore conclude that contact resistance is not the limiting factor, most importantly at the density values where the Lorenz anti-correlations associated with plasmon hopping are observed. Device 2, which has a measured resistance orders of magnitude larger than the quantum contact resistance, may possess a large extrinsic contact resistance occurring at the transition from graphene to the nanotube. This may play an important role in determining the absolute value of the Lorenz ratio. However, the clear gate dependence of the electrical and thermal conductances and the Lorenz ratio in this case suggests that other physics is at play, such as the plasmon hopping we propose.

## 12. NT Device 1 Data at Low Temperature

In this section, we present data from nanotube Device 1 at $T_{bath}$ = 5 K, shown in Fig.S9 below. This allows for a direct comparison with the corresponding bath temperature in Device 2. As such, the Device 2 data is reproduced here, using a wider gate voltage range to facilitate comparison.

In the Device 1 data taken at $T_{bath}$ = 5 K, the NT gap (charge neutrality point) shifted to a higher gate voltage as compared to the data in the main text. At this bath temperature, disordered Coulomb blockade is well-developed at lower density (higher gate voltage here). We observe correlation between the electrical and thermal conductances over the entire gate range, as seen previously. Anti-correlation

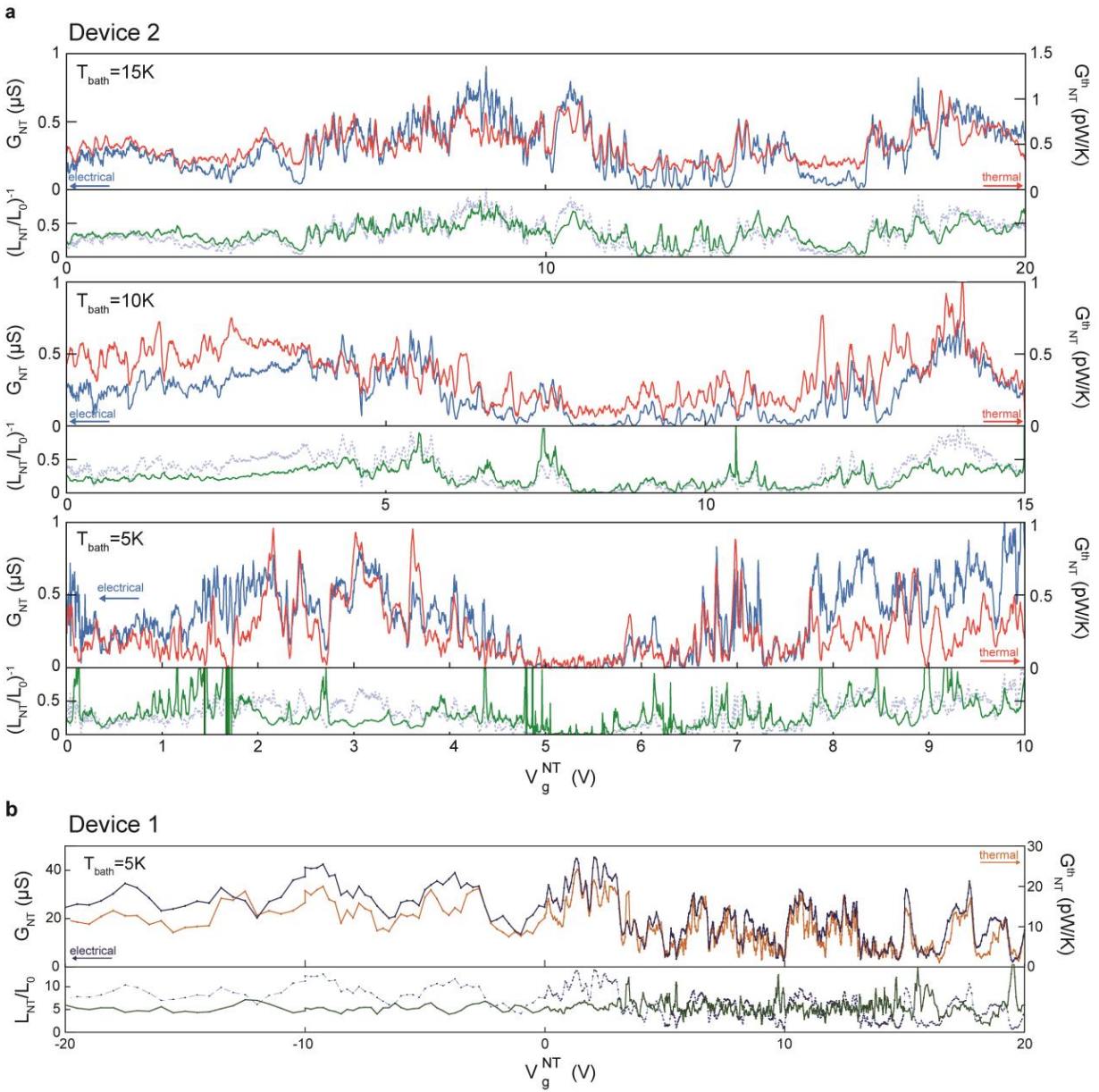

*Figure S9 Additional NT device data*. a) NT Device 2 data reproduced from Fig.3c of the main text, here with wider gate ranges. b) NT Device 1 measurement at $T_{bath} = 5K$. Top panel: blue, electrical conductance $G_{NT}$, orange thermal conductance $G_{NT}^{th}$. Bottom: green, Lorenz ratio $L_{NT}/L_0$, electrical conductance is reproduced as blue dashed line

between the Lorenz ratio and the electrical conductance is observed at the higher gate voltage range, where Coulomb blockade is more dominant. At the highest hole doping level (negative gate voltages), the Lorenz ratio becomes less gate dependent, suggesting that at these higher carrier densities the disorder potential landscape, which induces the barriers for plasmon hopping, is screened. As evidenced by the figure, we do not observe an indication of parasitic heat channels in this data. We note also that the Lorenz ratio exceeds 3 throughout this gate voltage range, including at low voltages where the Lorenz ratio is gate independent, consistent with contact contributions being small (see Supplementary Section 11).

## 13. Ruling Out Parasitic Contributions to Thermal Transport

In this section, we consider possible parasitic contributions to the measured thermal transport. First, we discuss the possibility of radiation between the hot and cold sides, including by plasmon hopping, and show that these are expected to be negligible. Next we discuss the possible effect of electron-phonon coupling in the nanotube bridges, and show that it is expected to be absent. Finally, we present a measurement of thermal transport in the nanotube gap in order to rule out parasitic contributions to within the measurement resolution.

Joule heating of the graphene thermometers, which raises their electron temperatures, may lead to enhanced emission of radiation. For the self-heating measurement of the hot and cold sides, radiative contributions have previously been shown to be negligible, leading to the observed Wiedemann-Franz regime observed in past works (see main text references 7, 23, 24, 34 and 37). However, we should consider the radiative coupling between the hot and cold sides. To quantify the effect, the geometric view factor should be considered. The geometric view factor gives a quantitative measure of the amount of thermal radiation leaving a source that arrives at some target body, taking into account their shapes, sizes, and relative distance and orientation. For the case of the 2D hot and cold sides in our device, the geometric view factor is that of two finite rectangles[23], which are also co-planar. Using the thickness of graphene (~0.3 $nm$) as an upper bound on the offset of the two, zero-thickness rectangles, we obtain a view factor of order $10^{-9}$. Thus the view factor in this geometry strongly suppresses radiative coupling.

Furthermore, since we operate at relatively low temperatures in this work, radiative contributions are suppressed by the $T^4$ power law of radiative emission. For instance, for a hot side of 5µm x 5µm, at T=5K, the total radiated power assuming a perfect blackbody emissivity is less than 1 fW, far below the Joule power in our typical measurements of order 1nW. To estimate the thermal conductance associated with radiative contributions, we consider a bath temperature of $T_0$ and a small thermal bias $\Delta T$, from which the thermal flux is $Q_{rad} \approx 4\,\sigma\,A\,F\,T_0^3 \Delta T$, where σ is the Stefan-Boltzmann constant, A is the cross-sectional area, and F is the geometric view factor computed above. Using the thickness of graphene and the device length, the radiative thermal conductance is then of order $10^{-18}$ pW/K at 1K, or $10^{-12}$ pW/K at 100K. Hence we expect that radiative contributions should contribute negligibly to the nonlocal thermal measurement. This is confirmed by the gate-independent Lorenz ratio equal to 1 in the graphene measurements of main text Fig.2, and the effective shut-off of thermal transport in the Coulomb and single-particle gaps of the nanotube devices in Fig.3.

The plasmon hopping mechanism of energy transfer that we consider in the main text could in principle also contribute to remote thermal coupling between hot and cold sides. However, such near-field mechanisms are ultimately dependent on a power law interaction between hot and cold ends. We considered Coulomb or screened Coulomb interaction, while other works discussing alternative mechanisms of heat transfer postulate van der Waals-type interactions[24–26]. The large power laws, ranging from 2-6, strongly suppress interaction across the large distance between hot and cold sides of 500 nm. Thus, we do not expect such near-field type mechanisms to be relevant for parasitic conduction between the graphene thermometers, as borne out by the data in Fig.S10, discussed below.

We now turn to consider the possibility of phonon contributions in the NT devices. Nanotubes have long been known to be electronically ballistic up to room temperature scales[27–32]. Our data shows a nanotube electrical conductance independent of temperature above the Coulomb blockade regime (see Fig.3b of the main text), consistent with the previously observed ballistic transport and with negligible electron-phonon coupling in the nanotube device.

Our thermal signals are also consistent with negligible electron-phonon coupling, as we now discuss. We first note that in the temperature range shown in the text ($T_{bath} \leq 70K$), electron-phonon coupling in graphene is weak, thus the thermal bias is largely electronic and excess energy injected by Joule heating is deposited in the electronic subsystem. Electron-phonon coupling in the NT bridge would cause some of this energy to transfer to the phonon subsystem of the bridge. However, the cold side thermometer, being also graphene, measures only the electronic part of the transported energy. Hence there would be *less* energy current transported to the electronic cold thermometer, resulting in a *lower* apparent thermal conductance. Thus, including electron-phonon energy transfer in the bridge will result in a *suppressed* Lorenz ratio, which is not observed. However, we instead observe *enhanced* Lorenz ratios, which are consistent with negligible electron-phonon coupling in the nanotubes, as previously found in the literature.

Energy transfer by electron-phonon coupling in the NT could still be possible by coupling NT phonons to graphene electrons. However, this higher order process involves either a remote coupling event (NT phonon to graphene electron) or two weak coupling events (NT phonon to graphene phonon and graphene phonon to graphene electron) and thus is highly improbable.

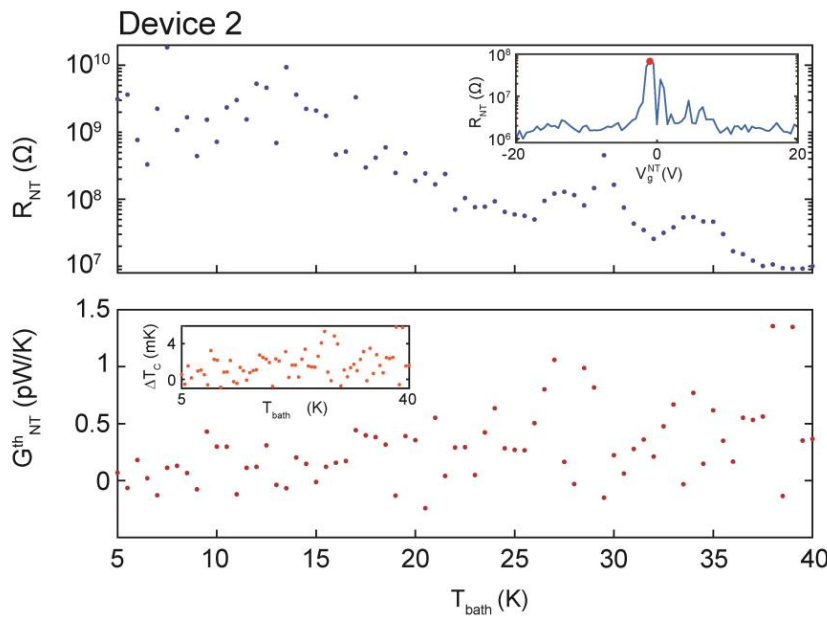

*Figure S10 Nanotube thermal conductance measurement in the bandgap.* Thermal bias $\Delta T_H/T_{bath} = 0.4$. Top panel: nanotube resistance $R_{NT}$ vs bath temperature $T_{bath}$. Inset: Nanotube resistance $R_{NT}$ vs gate voltage $V_g^{NT}$, red dot shows the gate set point in the bandgap, used in the other panels. Bottom: measured nanotube thermal conductance $G_{NT}^{th}$ vs $T_{bath}$. Inset: measured cold side temperature rise $\Delta T_C$ vs $T_{bath}$.

This picture may change significantly if the hot and cold graphene thermometers themselves possess electron-phonon coupling. This may happen, for instance, at higher bath temperature where the electron-phonon coupling in the graphene becomes non-negligible. In this case, the hot side can excite both the electrons and phonons of the bridge, while the cold side will measure energy transport associated with both the electrons and phonons. This regime is considered in Supplementary Section 14.

To confirm the absence of these parasitic energy transfer mechanisms, we turn to a measurement of NT Device 2, shown in Fig.S10. This device, with its larger gap, allows us to gate the NT into the single-particle gap, in which no excess charge is present and no electron transfer is allowed. We may thus completely shut off electrical conduction over a wider temperature range, such that only non-electrical contributions, such as the radiative or phonon mechanisms discussed above, may be present. We apply a large thermal bias $\Delta T_H/T_{bath} = 0.4$ and measure the nanotube thermal conductance $G_{NT}^{th}$ and cold side temperature rise $\Delta T_C$ while sweeping the bath temperature $T_{bath}$ and monitoring the DC resistance of the nanotube $R_{NT}$. Figure S10, top panel, shows the resulting $R_{NT}$ vs $T_{bath}$, which remains highly resistive over this temperature range but measurably decreases, from $\sim 10\ G\Omega$ to $\sim 100\ M\Omega$. Over the same temperature range, the measured $G_{NT}^{th}$ (Fig.S10, bottom) remains small enough to be essentially indistinguishable from zero at low temperatures, and as temperature increases it is nearly temperature independent, with significant scatter. The underlying cold side temperature rise $\Delta T_C$ (Fig.S10, bottom inset) remains close to the noise floor of $\sim 1 mK$ and is also nearly temperature independent. Over this temperature range, a radiative contribution with T[4] dependence would have shown an enhancement by a factor of more than 4000. In contrast, measurements of a bulk electrical insulator in a phonon-coupled regime show distinct contributions to thermal transport (see Supplementary Section 14). From this data, we conclude that no significant non-electronic energy transport channels are present in this temperature range, consistent with the above discussion.

## 14. Thermal Transport in an Electrical Insulator: α-RuCl$_3$

In the main text, we considered electronic thermal transport in 2D graphene and 1D carbon nanotubes, at low temperatures such that the graphene electron-phonon coupling of the thermometers is negligible. Three questions arise: i) At higher temperatures where the graphene electron-phonon coupling is non-negligible, can phonons contribute to the measured thermal transport? ii) Can we measure thicker samples in the bulk regime of behavior? iii) Are emergent neutral modes, such as magnetic modes, measurable with our technique? In this section, we present data from an H device in which the bridge is made of a microscale, bulk electrical insulator, the spin liquid candidate material α-RuCl$_3$, and show that the above questions are answered affirmatively.

The present H device is shown in Figure S11a. We utilize a 60 nm thick flake of α-RuCl$_3$ which becomes electrically insulating for T$_{bath}$ < 80 K. The device stack consists of a hBN bottom layer, two rectangular graphene sheets acting as noise thermometers, the α-RuCl$_3$, and a top encapsulating flake of hBN.

First, we characterize the hot side graphene thermal conductance versus temperature, shown in Fig.S11b, inset. As has previously been observed, the graphene follows electronic-diffusion-dominated thermal transport up until a bath temperature T* ≈ 44 K, above which electron-phonon coupling becomes non-negligible and contributes to energy loss (see main text references 7, 23, 24, 34 and 37). The transition temperature $T^*$ at which heat loss through electronic diffusion equals that lost through electron-phonon coupling is given by $T^* = \left(12L_0/\delta AR\Sigma_{ep}\right)^{1/\delta-2}$, where $R$ is the graphene two-point resistance, $A$ is the sample area, and $\Sigma_{ep}$ is the electron-phonon coupling constant, $\delta$ is the temperature exponent, and $L_0$ is the Lorenz constant[5]. We obtain $\Sigma_{ep} \sim 4.5 \times 10^{-4}\ W\ K^{-\delta}m^{-2}$, where we used $\delta = 4$, $T^* = 44K$, $A = 77\ \mu m^2$, and $R = 1100\Omega$. This value is larger than previous encapsulated graphene results. This may be due to the thick flake of α-RuCl₃, which possibly introduces additional channels of remote electron-phonon energy loss, since remote phonon coupling is believed to play a role in determining $\Sigma_{ep}$[7]. Here, the transition temperature is lower due to the increased area of the graphene hot side compared to past devices.

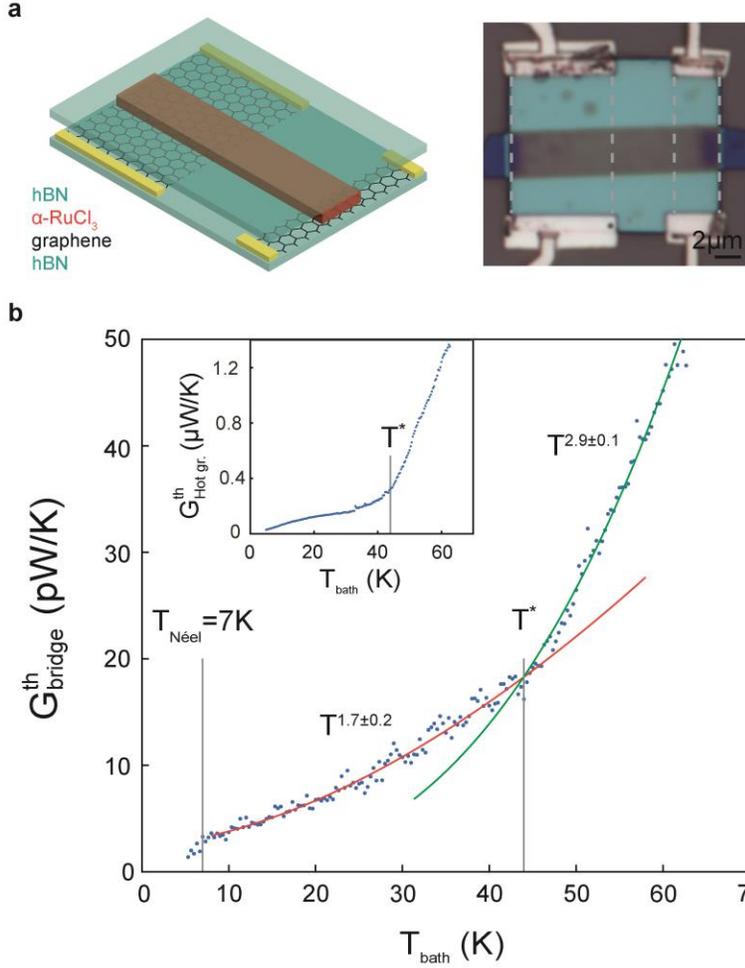

*Figure S11 Thermal measurement with nonlocal noise thermometry of α-RuCl₃. a) Left panel: Device material stack. Two rectangular graphene thermometers and the rectangular α-RuCl₃ bridge are encapsulated in hBN. Metal contacts are fabricated to the graphene. Right panel: optical micrograph of the device. Vertical dashed grey lines show boundaries of graphene hot (left) and cold (right) sides. α-RuCl₃ is the central dark horizontal strip. b) Bridge thermal conductance $G^{th}_{Bridge}$ versus bath temperature $T_{bath}$. Red curve: free-exponent fit to data between $T_{Neel}=7K$ and $T^*=44K$. Green curve: free-exponent fit to data between $T^*=44K$ and $T=63K$. Inset: Hot side graphene thermal conductance $G^{th}_{Hot\ gr.}$ versus $T_{bath}$ defined above.*

The bridge thermal conductance versus bath temperature, shown in the main panel of Fig.S11b, exhibits a distinct knee at T*. For T > T*, a free-exponent fit yields $G^{th}_{Bridge} \propto T^{2.9\pm0.1}$. The exponent value, close to 3, resembles that of bulk phonon thermal transport at low temperature. While this is a standard signature of bulk phonon thermal transport, it is notably different than past studies of bulk thermal transport in α-RuCl₃, which exhibit more complex behaviors[33,34]. One possible reason is the different size of sample, which here is few microns laterally and 60nm thick, as compared to typical hundreds of microns or mm lateral dimensions and dozens of micron thickness in bulk measurements. A second possibility is the measurement itself, which is performed with bulk metals as heaters and thermometers in traditional measurements, with different electron-phonon and electron-spin coupling properties.

For T < T*, despite the negligible electron-phonon coupling in the graphene heater, a signal is still observed. For this temperature range, the free-exponent fit yields $G_{Bridge}^{th} \propto T^{1.7\pm0.2}$. The abrupt change in exponent when the graphene heater decouples from phonons suggests a different heat transfer mechanism. At lower temperature, below 7 K, the thermal signal appears to show a rapid reduction. This hints at the known Neel transition temperature, below which α-RuCl$_3$ becomes antiferromagnetically ordered with gapped spin waves[34–36]. A rapid suppression of the signal is consistent with gapping out of the magnetic degrees of freedom below the transition. This raises the possibility that the measured heat transport in the intermediate temperature range is related to a magnetic degree of freedom. Past measurements have found evidence for a magnetic mode or continuum existing above the Neel transition[35,37]. In fact, one of the candidate theoretical models possesses linearly-dispersing 2D spinons[38], whose thermal transport exponent is expected to be 2. This behavior was observed in a different spin liquid candidate material[39]. Our preliminary observation is comparable to this expectation. However, the theoretical description is complicated by the presence of Heisenberg terms that give rise to the low-temperature antiferromagnetic state, and the true nature of these intermediate temperature degrees of freedom remains a topic of debate[40,41].

These results demonstrate that at high temperatures, when the graphene thermometers are electron-phonon coupled, phonon transport may be measured in the bridge. Combined with the analysis of Supplementary Section 6 and the low temperature, electron-dominated regime, we anticipate the ability to explore combined electronic and phononic transport in low-dimensional materials. These data also raise the possibility of studying thermal transport via magnetic degrees of freedom, including in insulating samples. Finally, this demonstration shows that this technique allows for bulk materials to be measured in a new regime of size, on the micron length scale compared to typical mm's required for bulk measurements.

## Supplementary References

1.  Talanov, A. V., Waissman, J., Taniguchi, T., Watanabe, K. & Kim, P. High-bandwidth, variable-resistance differential noise thermometry. *Rev. Sci. Instrum.* **92**, 014904 (2021).

2.  Sukhorukov, E. V. & Loss, D. Noise in multiterminal diffusive conductors: Universality, nonlocality, and exchange effects. *Phys. Rev. B* **59**, 13054–13066 (1999).

3.  Crossno, J. D. Electronic Thermal Conductance of Graphene via Electrical Noise. (2017).

4.  Pozderac, C. & Skinner, B. Relation between Johnson Noise and heating power in a two-terminal conductor. Preprint available at http://arxiv.org/abs/2104.05714 (2021).

5.  Fong, K. C. *et al.* Measurement of the Electronic Thermal Conductance Channels and Heat Capacity of Graphene at Low Temperature. *Phys. Rev. X* **3**, 041008 (2013).

6.  Betz, A. C. *et al.* Hot electron cooling by acoustic phonons in graphene. *Phys. Rev. Lett.* **109**, (2012).

7.  Crossno, J., Liu, X., Ohki, T. A., Kim, P. & Fong, K. C. Development of high frequency and wide bandwidth Johnson noise thermometry. *Appl. Phys. Lett.* **106**, 023121 (2015).

8.  Betz, A. C. *et al.* Supercollision cooling in undoped graphene. *Nat. Phys.* **9**, 109–112 (2012).

9.  Fong, K. C. & Schwab, K. C. Ultrasensitive and Wide-Bandwidth Thermal Measurements of Graphene at Low Temperatures. *Phys. Rev. X* **2**, 031006 (2012).

10. Crossno, J. *et al.* Observation of the Dirac fluid and the breakdown of the Wiedemann-Franz law in graphene. *Science* **351**, 1058–61 (2016).

11. Xue, J. *et al.* Scanning tunnelling microscopy and spectroscopy of ultra-flat graphene on hexagonal boron nitride. *Nat. Mater.* **10**, 282–285 (2011).

12. Thomsen, J. D. *et al.* Suppression of intrinsic roughness in encapsulated graphene. *Phys. Rev. B* **96**, 014101 (2017).

13. Zihlmann, S. *et al.* Out-of-plane corrugations in graphene based van der Waals heterostructures. *Phys. Rev. B* **102**, 195404 (2020).

14. Mao, J. *et al.* Evidence of flat bands and correlated states in buckled graphene superlattices. *Nature* **584**, 215–220 (2020).

15. Zhao, W. L. Z., Tikhonov, K. S. & Finkel'stein, A. M. Flexural phonons in supported graphene: from pinning to localization. *Sci. Rep.* **8**, 1–10 (2018).

16. Giamarchi, T. *Quantum physics in one dimension*. (Oxford University Press, 2003).

17. Kim, J., Ou, E., Sellan, D. P. & Shi, L. A four-probe thermal transport measurement method for nanostructures. *Rev. Sci. Instrum.* **86**, 044901 (2015).

18. Fong, K. C. Impact of contact resistance in Lorenz number measurements. Preprint available at https://arxiv.org/abs/1711.04005v2 (2017).

19. Chiatti, O. *et al.* Quantum thermal conductance of electrons in a one-dimensional wire. *Phys. Rev. Lett.* **97**, (2006).

20. Jezouin, S. *et al.* Quantum limit of heat flow across a single electronic channel. *Science (80-. ).* **342**, 601–604 (2013).

21. Banerjee, M. *et al.* Observed quantization of anyonic heat flow. *Nature* **545**, 75–79 (2017).

22. Srivastav, S. K. *et al.* Universal quantized thermal conductance in graphene. *Sci. Adv.* **5**, eaaw5798 (2019).

23. Howell, J. R. A Catalog of Radiation Heat Transfer Configuration Factors. Available at: http://www.thermalradiation.net/calc/sectionc/C-13.html.

24. Ong, Z. Y. & Pop, E. Molecular dynamics simulation of thermal boundary conductance between carbon nanotubes and SiO2. *Phys. Rev. B - Condens. Matter Mater. Phys.* **81**, 155408 (2010).

25. Persson, B. N. J., Volokitin, A. I. & Ueba, H. Phononic heat transfer across an interface: thermal boundary resistance. *J. Phys. Condens. Matter* **23**, 045009 (2011).

26. Pendry, J. B., Sasihithlu, K. & Craster, R. V. Phonon-assisted heat transfer between vacuum-separated surfaces. *Phys. Rev. B* **94**, 075414 (2016).

27. Frank, S., Poncharal, P., Wang, Z. L. & De Heer, W. A. Carbon nanotube quantum resistors. *Science (80-. ).* **280**, 1744–1746 (1998).

28. White, C. T. & Todorov, T. N. Carbon nanotubes as long ballistic conductors. *Nature* **393**, 240–241 (1998).

29. Poncharal, P., Berger, C., Yi, Y., Wang, Z. L. & De Heer, W. A. Room temperature ballistic conduction in carbon nanotubes. *J. Phys. Chem. B* **106**, 12104–12118 (2002).

30. Javey, A., Guo, J., Wang, Q., Lundstrom, M. & Dai, H. Ballistic carbon nanotube field-effect transistors. *Nature* **424**, 654–657 (2003).

31. Kong, J. *et al.* Quantum interference and ballistic transmission in nanotube electron waveguides. *Phys. Rev. Lett.* **87**, 106801 (2001).

32. Hertel, T. & Moos, G. Electron-phonon interaction in single-wall carbon nanotubes: A Time-domain study. *Phys. Rev. Lett.* **84**, 5002–5005 (2000).

33. Leahy, I. A. *et al.* Anomalous Thermal Conductivity and Magnetic Torque Response in the Honeycomb Magnet α-RuCl3. *Phys. Rev. Lett.* **118**, 187203 (2017).

34. Czajka, P. *et al.* Oscillations of the thermal conductivity in the spin-liquid state of α-RuCl3. *Nat. Phys.* (2021). doi:10.1038/s41567-021-01243-x

35. Banerjee, A. *et al.* Proximate Kitaev quantum spin liquid behaviour in a honeycomb magnet. *Nat. Mater.* **15**, 733–740 (2016).

36. Zheng, J. *et al.* Gapless Spin Excitations in the Field-Induced Quantum Spin Liquid Phase of α-RuCl3. *Phys. Rev. Lett.* **119**, 227208 (2017).

37. Sandilands, L. J., Tian, Y., Plumb, K. W., Kim, Y.-J. & Burch, K. S. Scattering Continuum and Possible Fractionalized Excitations in α-RuCl3. *Phys. Rev. Lett.* **114**, 147201 (2015).

38. Liu, Z. X. & Normand, B. Dirac and Chiral Quantum Spin Liquids on the Honeycomb Lattice in a Magnetic Field. *Phys. Rev. Lett.* **120**, 187201 (2018).

39. Yamashita, M. *et al.* Highly mobile gapless excitations in a two-dimensional candidate quantum spin liquid. *Science (80-. ).* **328**, 1246–1248 (2010).

40. Winter, S. M. *et al.* Breakdown of magnons in a strongly spin-orbital coupled magnet. *Nat. Commun.* **8**, 1152 (2017).

41. Laurell, P. & Okamoto, S. Dynamical and thermal magnetic properties of the Kitaev spin liquid candidate α-RuCl3. *npj Quantum Mater.* **5**, 1–10 (2020).